\documentclass{egpubl}
\usepackage{pg2021}
\JournalSubmission    

\usepackage[T1]{fontenc}
\usepackage{dfadobe}  

\usepackage{cite}  
\BibtexOrBiblatex
\electronicVersion
\PrintedOrElectronic
\ifpdf \usepackage[pdftex]{graphicx} \pdfcompresslevel=9
\else \usepackage[dvips]{graphicx} \fi

\usepackage{egweblnk} 
\usepackage{enumitem}
\usepackage{amsmath}
\usepackage{amsfonts}
\usepackage{multirow}
\usepackage{bigdelim}
\usepackage{subfigure}
\usepackage{color}
\usepackage{booktabs}
\usepackage{ifthen}
\usepackage{hyperref}
\usepackage{bbold}
\usepackage[noabbrev]{cleveref}
\usepackage{xcolor, pgf}
\usepackage{color, colortbl}
\usepackage[normalem]{ulem} 
\usepackage{placeins}
\usepackage{soul}
\usepackage[skins]{tcolorbox}

\usepackage{xr}
\makeatletter
\newcommand*{\addFileDependency}[1]{
  \typeout{(#1)}
  \@addtofilelist{#1}
  \IfFileExists{#1}{}{\typeout{No file #1.}}
}
\makeatother

\newcommand*{\myexternaldocument}[1]{%
    \externaldocument{#1}%
    \addFileDependency{#1.tex}%
    \addFileDependency{#1.aux}%
}
\myexternaldocument{supp}

\newlength\myboxwidth
\definecolor{gray}{rgb}{0.5,0.5,0.5}
\definecolor{green}{rgb}{0, 0.6, 0}
\definecolor{orange}{rgb}{1, 0.5, 0}
\definecolor{mahogany}{rgb}{0.75, 0.25, 0.0}
\definecolor{purple}{rgb}{0.6, 0, 0.6}
\definecolor{darkgreen}{rgb}{0, 0.3, 0}
\definecolor{orange}{rgb}{1, 0.5, 0.}
\definecolor{lightblue}{rgb}{0.52, 0.75,0.91}
\definecolor{softgreen}{rgb}{0.66,0.87,0.74}
\definecolor{softred}{rgb}{0.96,0.71,0.69}

\newcommand{\ignore}[1]{}
\newcommand{\none}[1]{}
\newcommand{\com}[1]{}

\newcommand{\etal}{{\textit{et~al.}}}
\newcommand{\ie}{i.e.,}
\newcommand{\eg}{e.g.,}

\DeclareMathOperator*{\argmax}{arg\,max}

\newcommand{\name}{EvIcon}
\definecolor{evalgray}{rgb}{0.4,0.4,0.4}
\definecolor{evalblue}{rgb}{0.38,0.612,1.0}
\definecolor{evalred}{rgb}{0.973, 0.463, 0.427}

\begin{document}
\title{EvIcon: Designing High-Usability Icon with Human-in-the-loop Exploration and IconCLIP}

\author[I-Chao Shen et al.]
{\parbox{\textwidth}{\centering I-Chao Shen$^{1}$ ~ Fu-Yin Cherng$^{2}$\thanks{Corresponding author} ~ Takeo Igarashi$^{3}$ ~ Wen-Chieh Lin$^{4}$ ~ Bing-Yu Chen$^{5}$
        }
        \\
{\parbox{\textwidth}{\centering $^1$ ichaoshen@g.ecc.u-tokyo.ac.jp
, The University of Tokyo, Japan \\
$^2$ fuyincherng@cs.ccu.edu.tw, National Chung Cheng University, Taiwan \\
$^3$ takeo@acm.org, The University of Tokyo, Japan \\
$^4$ wclin@cs.nycu.edu.tw, National Yang Ming Chiao Tung University, Taiwan \\
$^5$ robin@ntu.edu.tw, National Taiwan University, Taiwan \\
      }
}
}

\maketitle
\begin{abstract}
Interface icons are prevalent in various digital applications.
Due to limited time and budgets, many designers rely on informal evaluation, which often results in poor usability icons.
In this paper, we propose a unique human-in-the-loop framework that allows our target users, \ie~novice and professional UI designers, to improve the usability of interface icons efficiently.
We formulate several usability criteria into a perceptual usability function and enable users to iteratively revise an icon set with an interactive design tool, EvIcon.
We take a large-scale pre-trained joint image-text embedding (CLIP)
and fine-tune it to embed icon visuals with icon tags in the same embedding space (IconCLIP).
During the revision process, our design tool provides two types of instant perceptual usability feedback.
First, we provide perceptual usability feedback modeled by deep learning models trained on IconCLIP embeddings and crowdsourced perceptual ratings.
Second, we use the embedding space of IconCLIP to assist users in improving icons' visual distinguishability among icons within the user-prepared icon set.
To provide the perceptual prediction, we compiled \textit{IconCEPT10K}, the first large-scale dataset of perceptual usability ratings over $10,000$ interface icons, by conducting a crowdsourcing study.
We demonstrated that our framework could benefit UI designers' interface icon revision process with a wide range of professional experience.
Moreover, the interface icons designed using our framework achieved better semantic distance and familiarity, verified by an additional online user study.
\end{abstract}

\section{Introduction}
\label{sec:intro}
Amid the ubiquity of digital technologies, including computers, intelligent appliances, and wearable devices, interface icons play an increasingly important role in representing various functions with benefits including improving interface scannability (\ie~the ease of reading and understanding the content of the interface), saving space on small screens, and conveying information universally~\cite{setlur2005semanticons,setlur2014automatic}.
The usability of icons is determined by several characteristics, including visual complexity, style, familiarity, etc.~\cite{mcdougall2001effects,isherwood2007icon}
Existing design guidelines (\eg~Google's Material Design) provide designers with implications of icon design regarding visual characteristics.
However, collecting users' perceptual feedback on the icons is still an irreplaceable step to assess the icon's usability~\cite{bailey1992usability}.
Yet, conducting formal usability tests (\eg~inviting real users to perform usability tasks) can be time-consuming and requires extra effort~\cite{rosenholtz2011predictions,deka2017zipt,swearngin2019modeling}, which could significantly lengthen the iterative process of interface icon design.
Moreover, when evaluating icons designed for specific users (\eg~elders or users with lower computer literacy), conducting adequate usability testing is even more laborious.

\begin{figure}
    \centering
 \includegraphics[width=\linewidth]{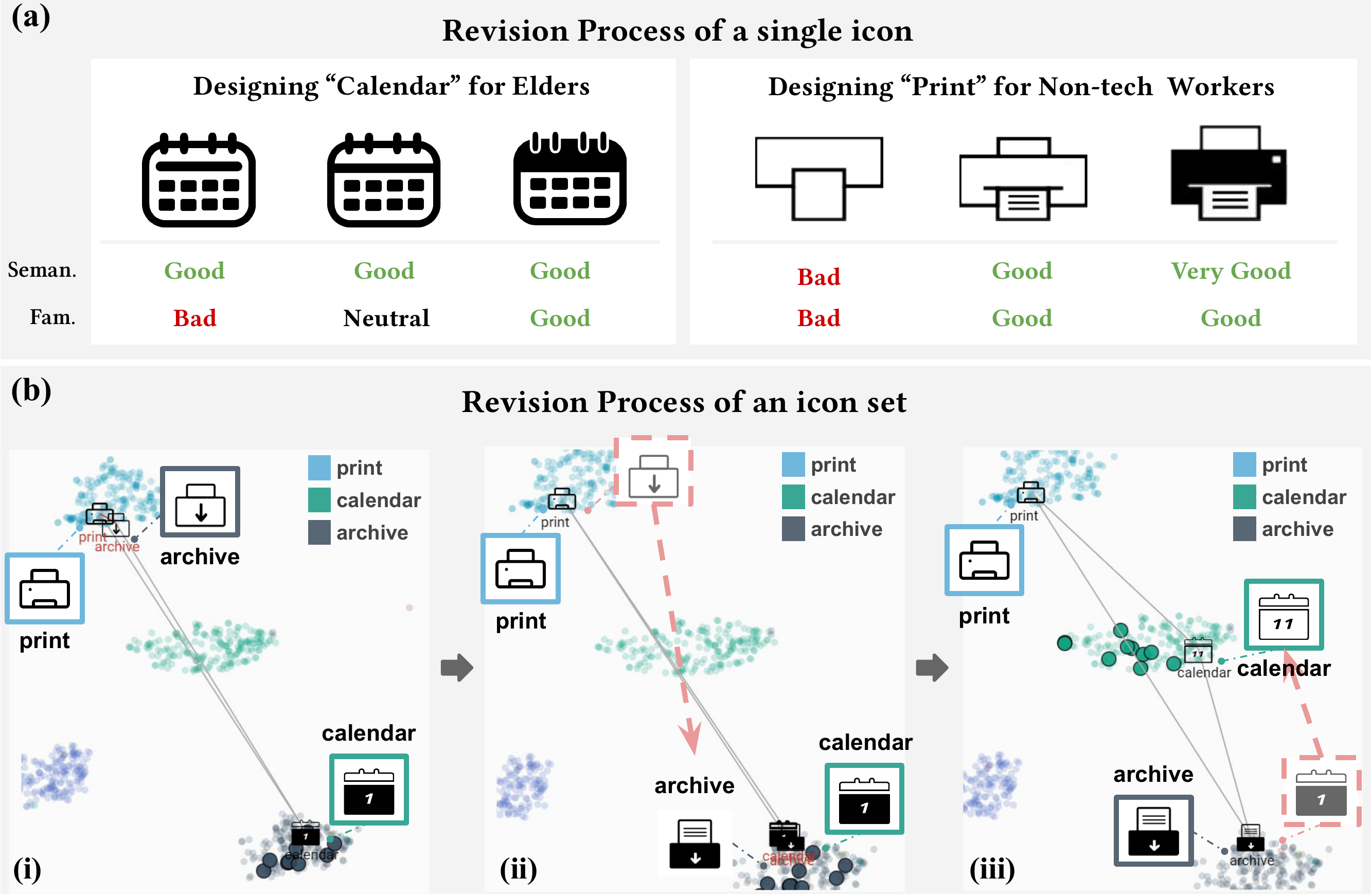}
 \caption{
 \name~provides two types of instant perceptual usability feedback.
 (a) An UI designer can improve a single icon's usability and target different demographic users (\eg~elder people or non-tech workers) with the ``semantic distance''  and ``familiarity'' feedback. 
 (b) Moreover, an UI designer can improve the usability of an icon set by (i) identifying poor visual distinguishability and (ii) revise the ``archive'' icon and (iii) the ``calendar'' icon using the visual distinguishability graph.
 }
 \label{fig:teaser}
\end{figure}
\begin{figure}[t!]
    \centering
    \includegraphics[width=\linewidth]{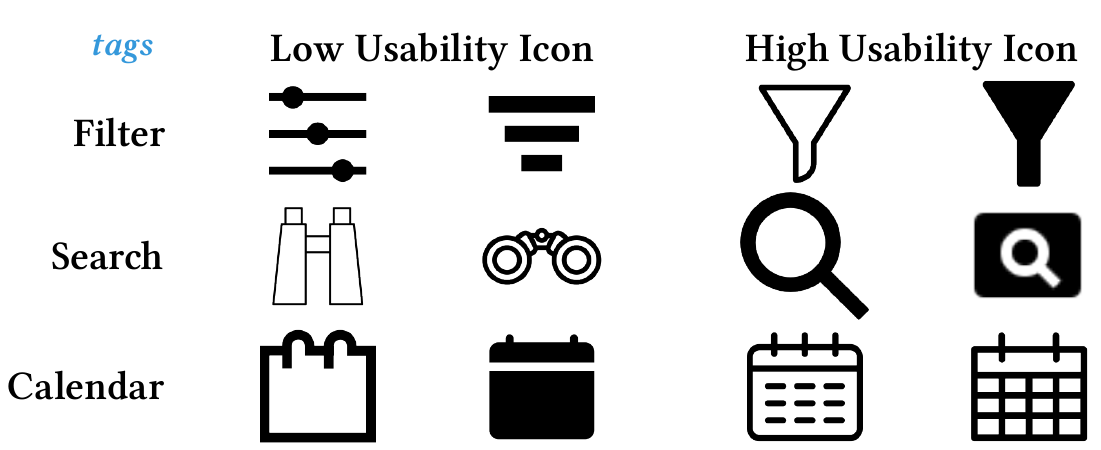}
    \caption{
    Example icons in \textit{IconCEPT10K} with usability rated by crowdworkers.
    Icons with low usability have three common shortcomings.
    First, these icons make users misunderstand their tag with others (\eg~Filter). 
    Second, they use unconventional metaphors to transmit the meanings of a concept (\eg~Search). 
    Last, these icons omit the critical features, so users fail to recognize the target concept (\eg~Calendar). 
    }
    \label{fig:bad_good_example}
\end{figure}

Zhao~\etal~\cite{zhao2020iconate} reported that designers often consult other UI designers' feedback on icons.
These informal evaluations often failed to provide comprehensive and objective information about how target users would perceive and use the icons~\cite{bailey1992usability,rosenholtz2011predictions}, thus leading to low usability icons. 
As shown in \mbox{\Cref{fig:bad_good_example}}, even for icons of standard tags (\eg~``Search'' and ``Calendar''), accidentally omitting the critical visual features when adjusting icons' style could lead to poor usability~\mbox{\cite{lin1994study,chajadi2020effects}}.
These examples further illustrate the importance of getting objective and instant feedback from users.
These findings underscore the need for an objective and comprehensive usability testing approach that is cost- and time-efficient.
Although several automatic graphical icon synthesis methods have been proposed~\cite{Lewis2004icons, KOLHOFF2008550}, involving humans (\ie~``human-in-the-loop'') in the design process has several advantages (\eg~solving computationally complex problems~\cite{holzinger2016interactive}, building users' trustworthiness to interactive systems~\cite{liao2020questioning}). 

Hence, instead of proposing an automatic icon synthesis method, we propose \name, an interactive framework to reduce the workload of performing usability tests for a user-prepared interface icon set.
\name~comprises two main parts: (i) a novel human-in-the-loop formulation of icon and icon set design and (ii) an interactive tool with instant perceptual usability feedback.
Our main idea is to formulate the common icon usability criteria into perceptual usability functions.
Among all icon-related features, we select \textit{semantic distance} and \textit{familiarity} as the usability criteria since they are the most critical indications of icons' effectiveness at conveying information~\cite{mcdougall1999measuring,setlur2005semanticons,warnock2013multiple,setlur2014automatic,cherng2016eeg} and are commonly used by professional artists.
\textit{Semantic distance} stands for the degree of closeness between an icon and the tag it represents and  
\textit{familiarity} referred to the frequency with which icons are encountered. 
In Figure~\ref{fig:teaser}(a), we show examples of icons with different semantic distances and familiarity levels.
Moreover, prior research has found that using icons with close semantic distance and high familiarity can significantly increase user performance on interfaces~\cite{mcdougall1999measuring,setlur2005semanticons,warnock2013multiple,setlur2014automatic,cherng2016eeg}. Hence, due to the importance of these two indications for icon's usability \cite{setlur2005semanticons, mcdougall2009s, setlur2014automatic}, we focus on providing icon designers with semantic distance and familiarity predictions on icon designs in this paper.
Moreover, as an icon is usually designed and displayed within an icon set~\cite{kurniawan2000rule}, we also use \textit{visual distinguishability} as a critical usability criterion for designing an icon set~\cite{kurniawan2000rule, Lewis2004icons, setlur2005semanticons}.
The goal is to prevent users from confusing icons of different tags.

To reach the goal of this study mentioned above, we gathered the first large-scale dataset of single-colored interface icon usability ratings coined as \textit{IconCEPT10K}.
The reason we focus on the single-colored icons is that single-colored icons have been recommended by popular online resources (\eg~Font Awesome and Noun Project) and major software providers (\eg~Google and Apple) due to their scalability in various screen sizes and applications as the prevalence of flat UI design \cite{spiliotopoulos2018comparative, legleiter2020flat}.
Also, icons are usually designed in single-colored in the first place and then edited their color later, tailoring to the configuration of display devices~\cite{goonetilleke2001effects, zhao2020iconate}.
Moreover, prior works found that icons' coloring is more critical to icons' visual attractiveness than effectiveness in the conveyance of meaning~\cite{hsieh2017multiple,chajadi2020effects,shen2021effects}. %
Accordingly, we consider devising icons in single-colored is common in the design process. 
Hence, we focus on the single-colored icons in the present study. 

Our perceptual usability function comprises two components. 
First, we took a large-scale pre-trained joint image-text embedding \mbox{(CLIP~\cite{radford2021learning})} and fine-tuned it to embed icon visuals with icon tags in the same embedding space (IconCLIP).
Second, we collected usability ratings for a curated icon dataset of $50$ base tags.
We expanded the base tags by using the tags associated with each icon; thus, our usability prediction model can recognize unseen tags and is scalable for future use. 
After building the perceptual usability function, we present an interactive user interface with two types of instant feedback (as shown in \Cref{fig:teaser}) to support refining icons' usability efficiently.
Users can iteratively revise the initial icon in the prepared icon set and query for predicted usability results.
The first feedback is the predicted perceptual usability of the revised icon (\Cref{fig:teaser}(a)).
The second feedback is the icon's visual distinguishability to other icons in (i) the user-prepared icon set and (ii) our icon dataset. 
This feedback is realized by providing an interactive two-dimensional visualization of the IconCLIP embedding (see~\Cref{fig:teaser}(b)).

To understand the benefits of \name~for designers, we conducted a user study with six UI designers and asked them to revise icon sets with and without using \name.
We further conducted an online user study on the revised icons to verify whether \name~can assist UI designers to improve icons' usability.
The result shows that \name~can assist UI designers with a wide range of professional experiences to improve the usability of their icon designs.
The major contributions and novelties of this paper include:
\begin{itemize}
    \item We propose a novel human-in-the-loop formulation, \name, for refining the usability of an icon set, while previous works focus on providing supports for designing a single icon ignoring the icon usability. 
    \item We gathered \textit{IconCEPT10K}, the first icon dataset with high-level perceptual usability ratings, instead of low-level visual perceptual properties such as visual saliency.
\end{itemize}
We implemented \mbox{\name} as a web application so anyone can test \mbox{\name} on their own icon set.
We will also release the source code, pretrained models, and the collected dataset (\textit{IconCEPT10K}).
\begin{figure*}[h!]
    \centering
    \includegraphics[width=\linewidth]{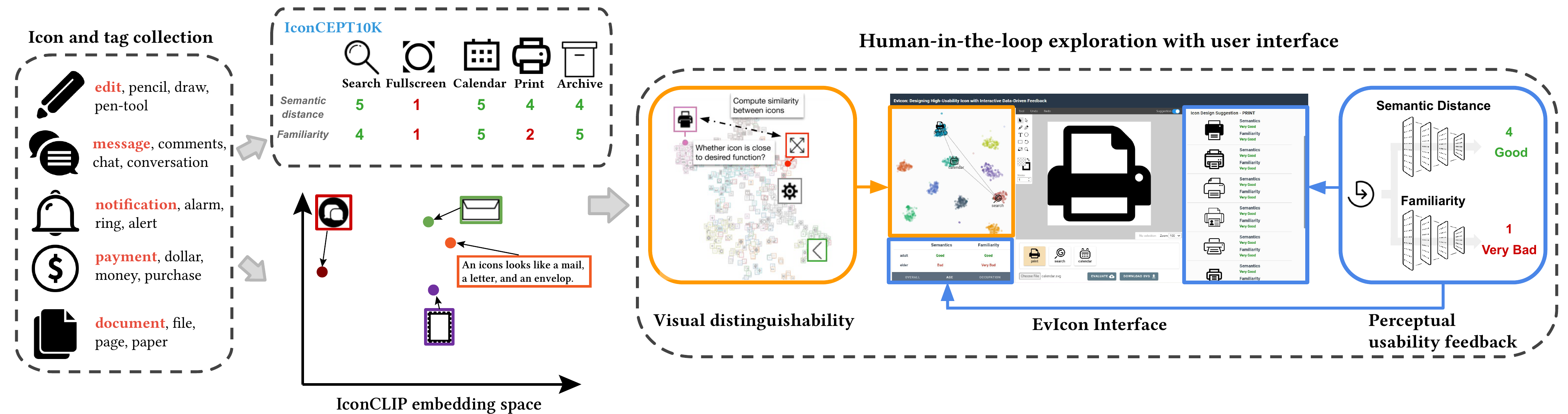}
    \caption{
    \textbf{Overview of \name.} 
    We collect a large-scale icon and tag collection. And we compiled a dataset \textit{IconCEPT10K}, comprises $10,000$ icons across $50$ base tags, their associated tags, and crowdsourced semantic distance and familiarity ratings.  
    We also fine-tune a pre-trained joint text-image embedding (CLIP) into IconCLIP using this collection.
    \name~computes and presents designers with instant perceptual usability feedback to assist revising high-usability icon sets.
    }
    \label{fig:evicon_overview}
\end{figure*}
\section{Related Works}
\subsection{Icon Design and Analysis}
Icon plays an essential role in visual communication, including graphic design and user interface design.
Prior studies~\cite{gittins1986icon,Horton1994,Horton1996} provide a thorough introduction on how to design icons and recommended practices.
Icon's usability is mainly associated with the ability to convey the information it represents. 
Previous research identified several features that heavily influence an icon's usability, including visual complexity, semantic distance, and familiarity \cite{mcdougall2001effects, mcdougall2009s,setlur2014automatic,kamarulzaman2020comparative}. 
Some studies reported that the users' age~\cite{leung2011age} and experience~\cite{isherwood2007icon,ali2021anachronism} also influence the effect of these features on icons' usability.

Researchers have proposed various methods to support icon design and generation due to the complex relationship between icons' features and usability. 
Zhao~\etal~\cite{zhao2020iconate} developed a system to generate icons containing compound meanings automatically.
Some works focus on generating icons based on file-names~\cite{Lewis2004icons}, data content~\cite{KOLHOFF2008550}, and man-made object category~\cite{shen2021clipgen}.
Other prior works focus on learning icons' appearance similarity~\cite{lagunas2018learning}, creating scale variations of icons~\cite{BL:2015:lillicon}, and selecting an icon set based on crowdsourced ratings~\cite{laursen2016icon}.
Compared to previous works~\cite{laursen2016icon,lagunas2018learning}, our system lets designers devise the final icon set on their own with our perceptual usability feedback instead of directly obtaining an icon set from an optimization process. 

\subsection{Assistive Authoring Tool for Visual Design}
Assistive visual content authoring has gained increasing interest in the past few years since the surge of the need for novel visual content.
Many works utilized personal editing histories to assist 2D sketch~\cite{Xing:2014:APR}, 3D shape sculpturing~\cite{Peng:2018:AS}, and viewpoint selection~\cite{Chen:2014:HAV}.
On the other hand, various prior works have incorporated real-time physical simulation into their interactive tools for designing physically valid furnitures~\cite{umetani2012guided} and model airplanes~\cite{umetani2014airplanes}.
Among them, many recent works leveraged collected visual content data to assist 2D sketch~\cite{Lee:2011:SRU}, multi-view clipart design~\cite{shen2020clipflip}, and mobile apps user interface design~\cite{liu2018learning, Rico, ZIPT}.
Other studies crowdsourced and modeled large-scale users' perception about tappability for the mobile interfaces~\cite{swearngin2019modeling} and visual importance on graphic designs~\cite{bylinskii2017learning} to assist designers in diagnosing the perceptual issues in their designs.
Additionally, Rosenholtz~\etal~\cite{rosenholtz2011predictions} conducted a thorough qualitative study with professional design teams and showed that designers benefited from tools with low-level perceptual prediction in the agile assessment of usability.
Unlike previous works that only focused on providing low-level visual perceptions feedback, we provide high-level usability feedback such as semantic distance and familiarity.
Moreover, we provide visual distinguishability feedback to support revising an icon set's usability, which is rarely addressed in prior related research.

\subsection{Human-in-the-loop Exploration}
Prior studies have demonstrated the feasibility of conducting usability evaluation on crowdsourcing platforms via performing benchmark user testings~\cite{komarov2013crowdsourcing} and collecting human visual importance \cite{bylinskii2017learning}.
As exploring various huge design spaces with usability evaluations is a ubiquitous task in visual design, this task is realized by various interactive optimization techniques, including interactive evolutionary computation~\cite{takagi2001interactive} and human-in-the-loop Bayesian optimization~\cite{koyama2014crowd, yuki2017sls,koyama2020sequential,chong2021interactive,brochu2010bayesian}.
Unlike previous methods, our human-in-the-loop framework focuses on providing instant perceptual usability feedback to support users' exploration instead of providing the final design using the optimization-based method due to the following reasons. 
First, the state-of-the-art human-in-the-loop optimization methods work best in relatively lower-dimensional parameter spaces (\eg~$6-15$) \cite{koyama2014crowd, yuki2017sls,koyama2020sequential,chong2021interactive,brochu2010bayesian}, whereas reducing the design dimensions of interface icons into such low dimensions would omit the nuanced features that are crucial for the high-level usability perceived by users and designers.
Hence, our framework makes designers finalize the icons and the icon sets iteratively and manually.
Second, previous human-in-the-loop optimization methods use ``selection'' as the main interaction approach, whereas the task of icon design requires more complicated design interactions than selections \cite{zhao2020iconate}.
Therefore, the current human-in-the-loop optimization methods are not suitable for the inputs of our framework to design high-usability icons.
\section{Problem Overview}
\label{sec:problem_overview}
Given an interface icon set $\mathcal{I}$ provided by a designer. 
The goal of our framework is to assist this designer in revising the usability of prepared icons into a new interface icon set $\mathcal{\hat{I}}$ efficiently.
We expect that each icon $I$ in $\mathcal{I}$ is associated with $n$ text tags ($\mathcal{T}_{I}={t_0, t_1,...,t_{n-1}}$) that represent the semantic and visual concepts of the icon such as ``search'', ``next'', ``television'', and ``map''.
We characterize the usability of an icon using common perceptual usability metrics including \textit{semantic distance}, \textit{familiarity}, and \textit{visual distinuguishability}, which are commonly used by professional icon designers~\cite{mcdougall2001effects, kurniawan2000rule,setlur2014automatic}.
However, these metrics of an icon are usually hard to evaluate mathematically from the icon image since the assessments of these metrics require extensive user testing to collect users' self-reports and feedback.
Hence, we collected a large-scale icon dataset and the crowdsourced perceptual ratings of these icons on Amazon Mechanical Turk (AMT). 
We used the collected ratings to train usability classifiers.
For each tag $A$, we trained a separate classifier $f^{\text{sd}}_{A}$ and $f^{\text{fam}}_{A}$ for classifying the semantic distance and familiarity of an icon belongs to tag $A$.
For each classifier, it predicts ``Very Good'', ``Good'', ``Neutral'', ``Bad'', and ``Very Bad'' as the different levels for the semantic distance and familiarity.
The goal of our framework is to enable users to revise an icon $I\in\mathcal{I}$ that maximizes the following perceptual usability function:
\begin{align}
    {i}^* = \argmax (w_{sd}\phi_{sd}(I,\mathcal{T}_I)+w_{fam}\phi_{fam}(I,\mathcal{T}_I) + w_{vd}\phi_{vd}(I)),
\label{eq:usability_func}
\end{align}
where $\phi$ is a semantic perceptual function. 
In our work, the semantic perceptual function comprises three parts:
\begin{itemize}
    \item semantic distance: $\phi_{sd}(I,\mathcal{T}_I) =  P(f^{\text{sd}}_{A} ==\text{Very Close}|I,\mathcal{T}_I)$
    \item familiarity: $\phi_{fam}(I,\mathcal{T}_I) =  P(f^{\text{fam}}_{A} ==\text{Very Good}|I,\mathcal{T}_I)$
    \item visual distinguishability: $\phi_{vd}(I) = \sum_{J\in\mathcal{I}} \|\rho_{I}-\rho_{J}\|_2^2$,
\end{itemize}
where $P(f^{\text{sd}}_{A} ==\text{Very Close})$ stands for the probability of an icon being classified as having the ``Very Close'' semantic distance.

To measure visual distinguishability, it is important to measure the distance with respect to the semantic concept difference instead of just pixel-level difference.
To achieve this, we obtain an embedding space where icons of the same tags stay closer to each other than those of different tags. We describe how to obtain this embedding space in \Cref{sec:icon_clip}.
The embedded coordinates of icon $i$ in this space are represented by $\rho_i$. The goal of $\phi_{sd}(i)$ is to encourage the revised icon to be classified as ``Very Close,'' while the aim of $\phi_{vd}(i)$ is to separate the refined icon from other icons in $\mathcal{I}$. To optimize \Cref{eq:usability_func} and iteratively refine the icon set $\mathcal{I}$, designers need to be involved in the process to specify their design requirements. Instead of providing designers with automatic synthesis results, we have developed an interactive interface that guides them in designing highly usable icons.

\section{EvIcon User Interface}
We propose an interactive and exploratory design tool, \name, to present perceptual usability feedback of an individual icon and visual distinguishability between icons.
Our interface augments existing vector graphics design tools with additional usability feedback panels.
As shown in Figure \ref{fig:evicon_overview}, our interface contains three main panels:
(i) the main canvas panel which includes a vector graphics editor for icon revision and a list to present the uploaded icon set,
(ii) the perceptual feedback panel (box with blue borderline), and
(iii) the distinguishability visualization panel (box with orange borderline).

\subsection{User Workflow}
To use \name, a designer first prepares a set of icons and corresponding tags under designing.
Next, the designer can select an icon from the icon set, and \name~would infer its predicted usability.
Then, the designer can check the predicted usability for general users or users with particular demographics in the perceptual usability feedback panel.
Furthermore, the designer can revise the selected icon to improve its usability and inspect the visual distinguishability of the revised icon using the interactive distinguishability graph. 
The designer can repeat these steps until the perceptual usability of the selected icon or the visual distinguishability between icons is satisfied.

\subsection{Interface Components}
\begin{figure}[t!]
    \centering
    \subfigure[Initial icon.]
    {\label{fig:bad_adjust:1}\includegraphics[width=0.3\linewidth]{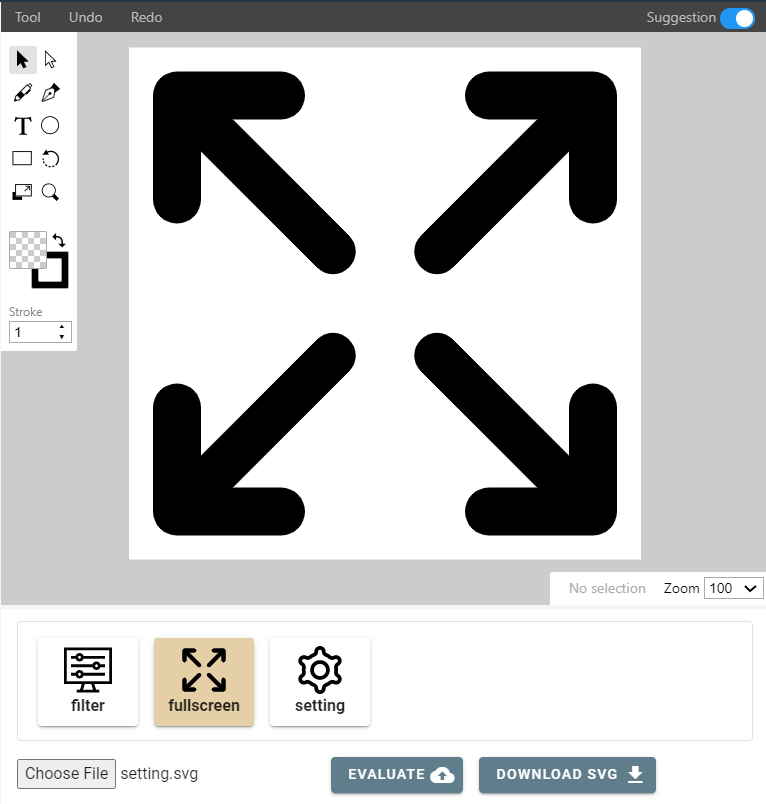}}
    \subfigure[Modified icon.]
    {\label{fig:bad_adjust:2}\includegraphics[width=0.3\linewidth]{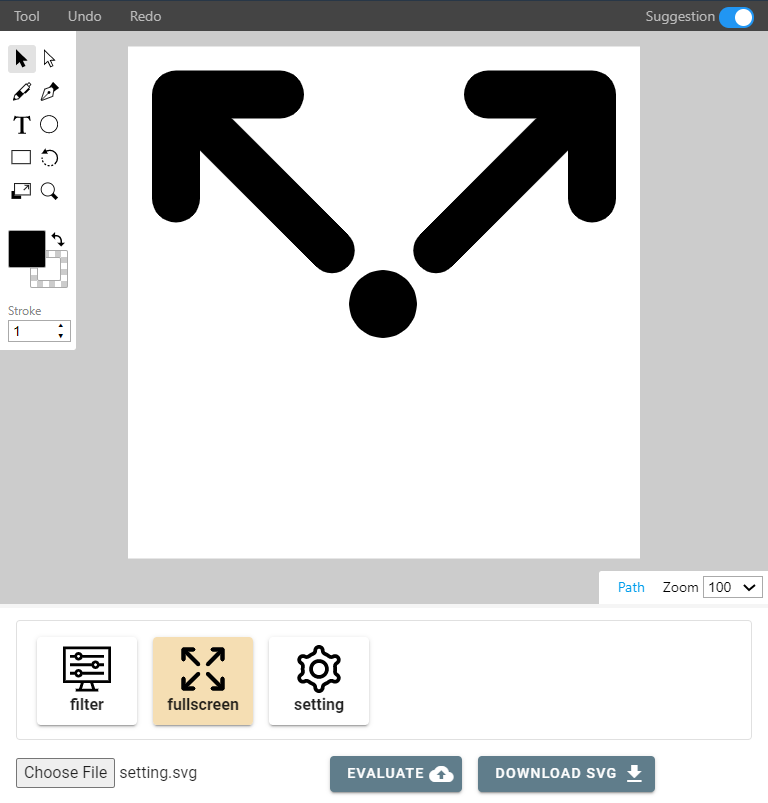}}
    \subfigure[Two types of in-place warnings.]
    {\label{fig:bad_adjust:3}\includegraphics[width=0.3\linewidth]{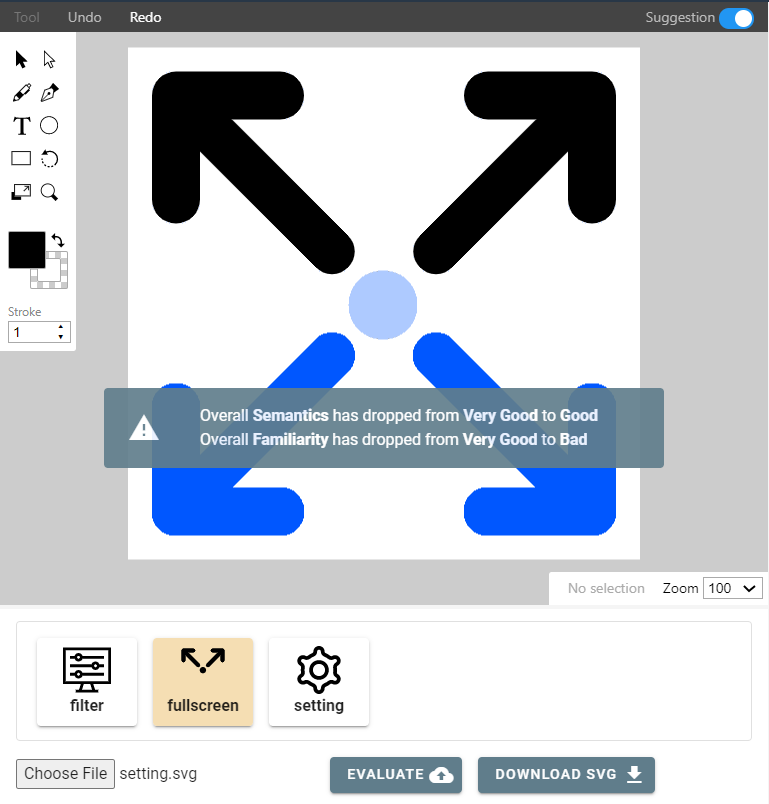}}
    \caption{
    A warning will be displayed in place to draw attention to the poor adjustment compared to the last usability inspection. Blue highlights will indicate the paths suggested to be added back, while light-blue highlights will mark those suggested to be removed.
    }

    \label{fig:bad_adjust}
\end{figure}
\subsubsection{Main Canvas Panel}
The designer can revise the icons using the vector graphics editor in this panel.
During the iterative revision process, \name~also provides in-place warnings when the predicted perception usability drops.
This in-place visual warning is helpful for building the connection between the revised icon and the perceptual prediction.
We highlighted the paths of an icon that we encourage the designers to add and remove in two different colors as shown in \Cref{fig:bad_adjust}.
\begin{figure}
\centering
    \subfigure[Overall]
    {\label{fig:pred_all}\includegraphics[scale=0.25]{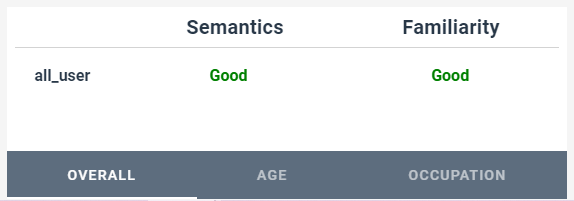}}
    \hfill%
    \subfigure[Age categories: i) adult and ii) elder]
    {\label{fig:pred_age}\includegraphics[scale=0.25]{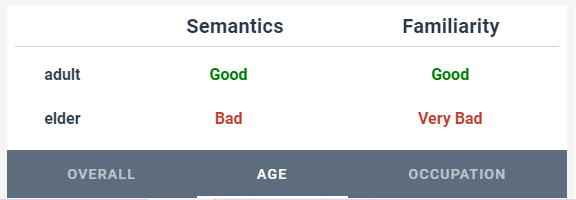}}
    \hfill%
    \subfigure[Occupation categories: i) technology, ii) business, and iii) others]
    {\label{fig:pred_occ}\includegraphics[scale=0.25]{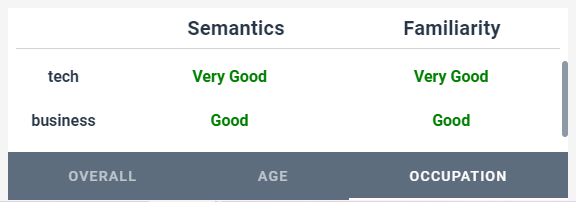}}
    \caption{\name~ provides predicted perceptual usability feedback. Apart from viewing perception feedback for general people (a), users can inspect the feedback from different demographics categories including (b) age and (c) occupation.
    }
    \label{fig:pred_panel}
\end{figure}
\subsubsection{Perceptual Usability Feedback Panel}
\name~shows the predicted level of perceptual usabilities (semantic distance and familiarity) of the icon under revision (\Cref{fig:pred_panel}).
The designers can switch between tabs to check the predicted usabilities for target audiences with particular demographics. 
To present the levels of semantic distance and familiarity in a way designers can easily understand, instead of showing rating scores directly, we use ``Very Bad'', ``Bad'', ``Neutral'', ``Good'', and ``Very Good'' to represent five different levels of user perceptions, and semantic distance is presented as ``Semantics'' on the interface of \mbox{\name}.
We highlighted ``Very Bad`` and ``Bad'' in red, ``Neutral'' in black, and ``Good'' and ``Very Good'' in green to enhance readability.

\subsubsection{Distinguishability Visualization Panel}
\label{sec:interface_viz}
\begin{figure}[h!]
\centering
  \includegraphics[width=1.0\linewidth]{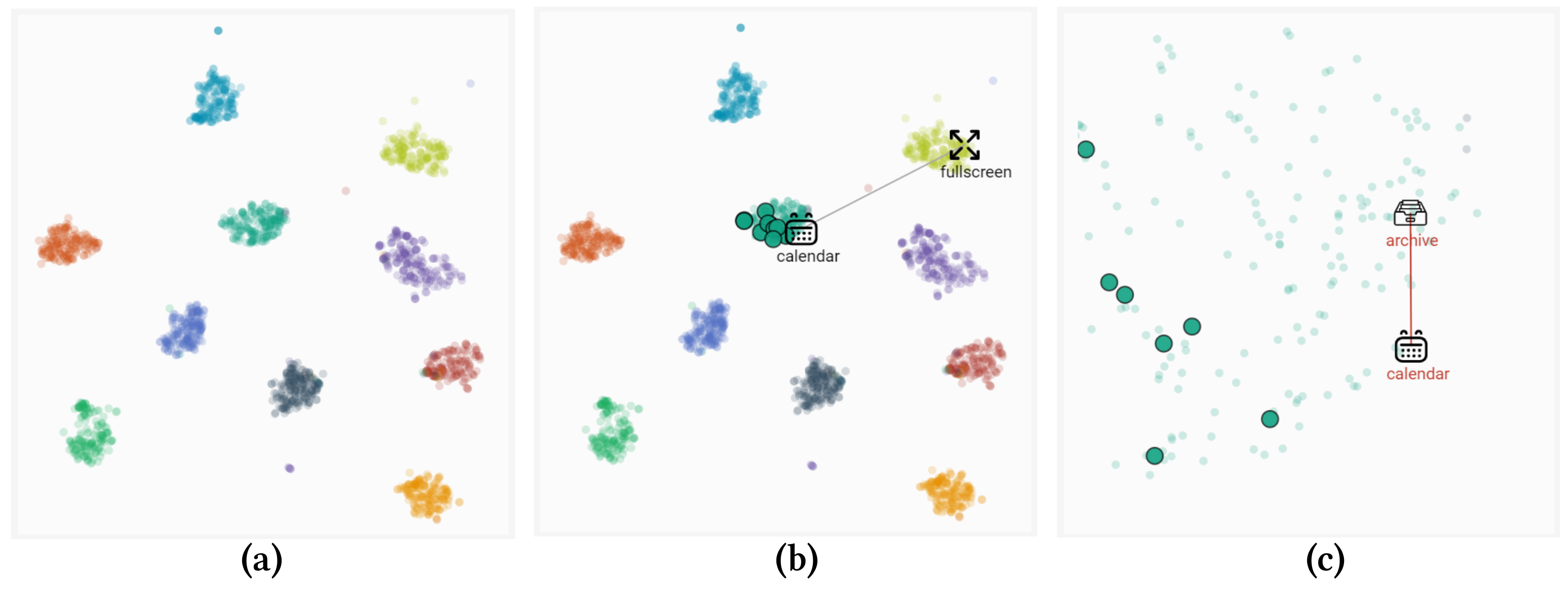}
  \vspace{-0.5cm}
  \caption{
(a) Different color-codings indicate different semantic concept clusters.
The icons are linked (b) in grey but will change (c) to red in order to notify poor visual distinguishability.
    }

\label{fig:dgraph}
\end{figure}

\name~presents an interactive distinguishability graph to help designers compare the relative visual distance between icons in the prepared icon set $\mathcal{I}$.
After the designers revise an icon, they can check the updated embedded coordinate of the icon.
We connected the icons in the prepared icon set using grey links (as shown in \Cref{fig:dgraph}(b)) and changed the color of the links into red if the connected icons were too close to each other (see~\Cref{fig:dgraph}(c)).
This interactive design aims to warn designers of the inadequate visual distinguishability in the prepared icon set, and prevent them from refining icons that fall into the wrong tag.

\section{EvIcon Implementation}
\label{sec:impl}
\subsection{Icon and Crowdsourced Perceptual Rating Dataset}
\label{sec:dataset}
Our goal of data collection is to gather an icon dataset covering a comprehensive range of tags that UI designers are likely to design.
Since unlimited tags exist for interface icons, it is impractical to enumerate them all and collect them at once.
To address this issue, we expanded the tags we can cover by adopting the following data collection procedure.
First, we collected icons of $212$ base tags reported in prior work~\cite{liu2018learning}, including ``Search'', ``Crop'', ``Message'', ``Pause'', ``Filter'', ``Calendar'', and ``Archive''.
We collected these single-colored icons from multiple online resources, including Google Material Icons, Icon8, and The Noun Project.
Although these icons are collected from different websites, they share similar visual styles due to the prevalence of flat UI design \cite{arledge2014filled, spiliotopoulos2018comparative}. 
Overall, we collected $2,613,438$ single-colored icons and their associated tags provided by the original designers.
There are $191,472$ unique tags representing a wide range of concepts, and they provided us with a rich resource to model the relationship between icons and tags.
We then used this icon and tag collection to train a joint image-text embedding. 

However, it is tedious and repetitive to collect users' perceptual usability ratings for all icons; thus we selected the top $50$ base tags that are semantically independent based on the icon numbers of each tag.
For each selected base tag, we further selected $200$ representative icons with respect to the uniqueness of icon shapes using the following process.
After normalizing the size of icons from different resources into $28\times 28$ pixels, we applied the principal component analysis (PCA) on icons' pixel values after removing the duplicated icons.
Then, we set the projection to preserve $90\%$ of the variances to generate the final principal components and utilize them to represent each icon.
Next, we performed K-Means clustering~\cite{kmeans} on these projected icon representations and set $K=10$ based on the results of the Elbow method (\ie~ten clusters in a subset) \cite{ketchen1996application}.
We obtained $200$ icons from each base tag by randomly sampling $20$ icons from each cluster. 
After repeating the same process to all base tags, we acquired the curated dataset with in total $10,000$ icons in which the variety of icons of each function increased compared to the raw dataset.

After obtaining the curated dataset, we used Amazon Mechanical Turk (AMT) to collect users' perceived semantic distance and familiarity with the selected $10,000$ icons.
We recruited $5,559$ workers participating in the crowdsourcing task (3,498 males and 2,061 females; mean age = 33.1 with a standard deviation of 8.90).
The workers' self-report ages and occupations were divided into three age levels (elder: age $>50$ yrs; adult: $50>$ age $>20$ yrs; teenager: age $<20$ yrs) and occupational categories (technology, business, and others) which are used as the demographic information of their ratings when building perceptual usability prediction.
Each worker finished five assignments and rated icons of five tags in each assignment (\ie~$25$ icons in total) with an average completion time of $8$ minutes. The workers were asked to rate each icon on a 5-point Likert scale to specify their assessment of the icon’s semantic distance and familiarity \cite{mcdougall1999measuring,isherwood2007icon}. 
The workers also rated their perceived familiarity with each tag on the same 5-point Likert scale. 
We described the rating distribution of the $50$ base tags and the content of the questions in \Cref{sec:supp_amt_collection} of the supplementary material.
The order of the icons was randomized.
In general, we spent two days collecting all the rating data in parallel using MTurk API.
In the final rated dataset, we collected $138,964$ unique ratings.
We describe the details of the distribution of the collected ratings and the AMT crowdsource task in \Cref{fig:per_dist} of the supplemental material.
We will include the selected $10,000$ icons and the collected perceptual usability ratings as our \textit{IconCEPT10K} dataset.
\subsection{Perceptual Usability and Visual Distinguishability Feedback}
\label{sec:icon_clip}
Given an input icon $I$ and its associated tags $\mathbf{T}={t_0, t_1,...,t_{n-1}}$, we want to build a classifier that can predict its perceptual usability ratings (semantic distance and familiarity).
However, there are unlimited possible tags designers want to design; and it is tedious to collect icons of all possible tags and their perceptual usability ratings.
Thus, it is vital to design a classification method to predict the perceptual usability ratings for icons of unseen tags.
To address this need, we designed our classification method based on the pre-trained joint embedding (CLIP)~\mbox{\cite{radford2021learning}} which is learned from loose image-text pairing information.
\subsubsection{Introduction to CLIP Embedding Space}
CLIP~\mbox{\cite{radford2021learning}} is a joint image-text embedding trained on $400$ million text-image pairs.
The representations learned by CLIP have been shown to be effective for various downstream tasks such as zero-shot image classification.
CLIP jointly trains an image encoder $g$ and a text encoder $h$, that map images and text into a shared embedding space.
Unlike previous works on natural image editing using CLIP embedding space~\mbox{\cite{Patashnik_2021_ICCV, abdal2021clip2stylegan}}, the target image domain of our application (single-colored icon image) is different from the training images used in the pre-trained CLIP model.
Thus, instead of using the pre-trained CLIP model to extract image and text representations directly, we finetune the original CLIP model using our icon dataset to obtain IconCLIP.
\begin{description}[style=unboxed,leftmargin=0cm]
\item[Finetuning CLIP on icon image]
We let $S_{\text{icon}}=\{(I_i,\mathbf{T}_i)|i=0,...,N\}$ denote the icon dataset used for finetuning the original CLIP model.
For each icon $I_i$, we converted the associated tags $\mathbf{T}_i$ into a sentence $s_i$ using the prompt template ``A icon looks like a \{$\textit{tag}_0$, $\textit{tag}_1$, ..., $\textit{tag}_{n-1}$\}''.
We use a pre-trained CLIP ViT-B/32 model as the base model, which uses ViT-B/32~\mbox{\cite{dosovitskiy2020vit}} as the image encoder and Transformer~\mbox{\cite{Vaswani2017trans}} as the text encoder.
We follow the training procedure described in the original CLIP~\mbox{\cite{radford2021learning}}.
Given a training pair (an icon image $I_i$ and a sentence $s_i$), CLIP produces a scalar score: $g(I)^T h(s_i)$ that is high when the image and text are mismatched.
We finetune the pre-trained model by minimizing a symmetric InfoNCE loss~\mbox{\cite{van2018representation}}. 
\end{description}
\subsubsection{Perceptual Usability Prediction}
\label{sec:perp_predict}
\begin{figure}[t!]
\centering
  \includegraphics[width=1.0\linewidth]{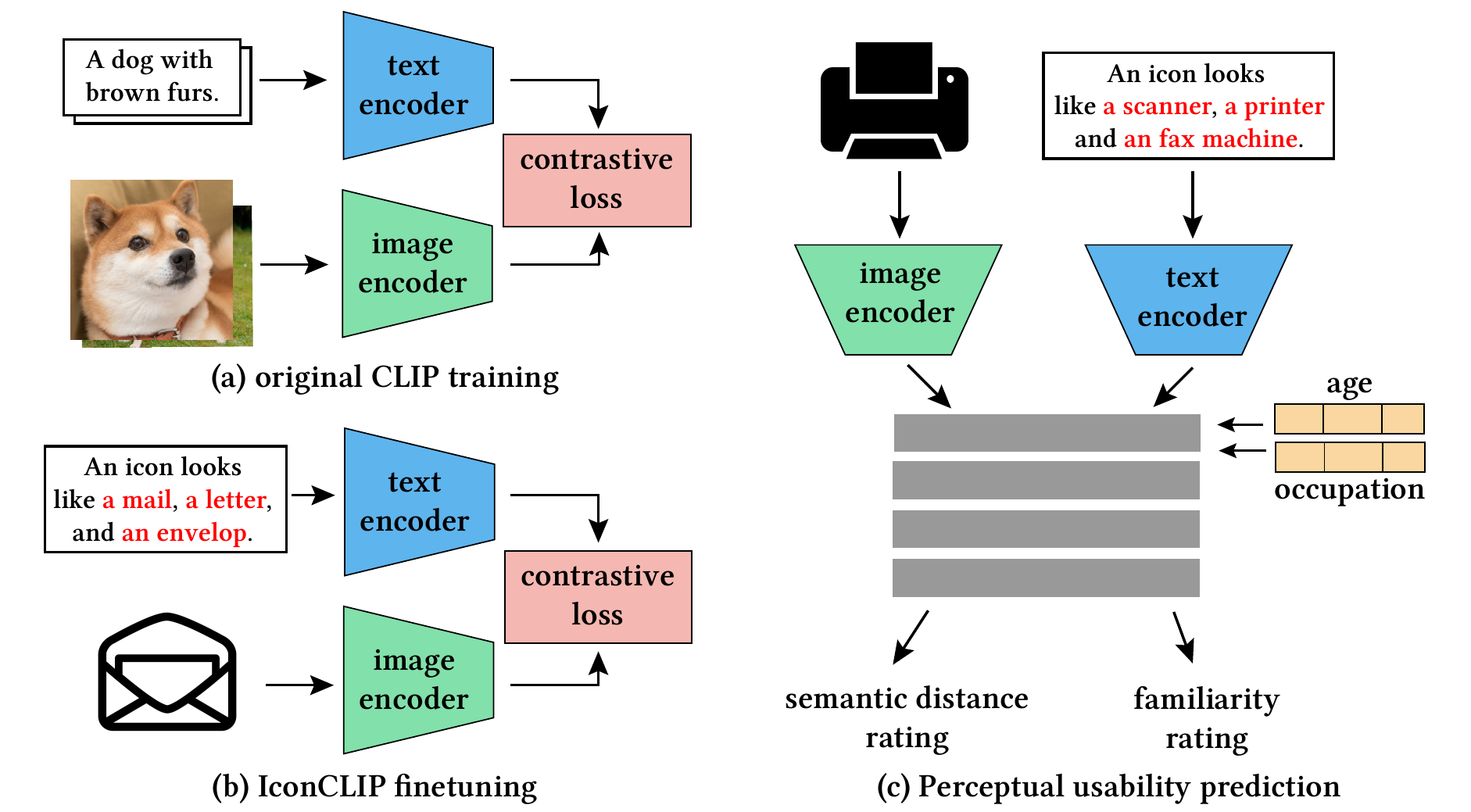}
  \caption{
    (a) The general-purpose CLIP~\cite{radford2021learning} is a joint image-text embedding trained on $400$ million text-image pairs.
    (b) We fine-tune the general-purpose CLIP into IconCLIP using ``icon tags''-``icon image'' pairs.
    (c) We predicted the perceptual usability ratings of an input icon and its associated tags using the IconCLIP embedding space and the target demographic information.
    }
\label{fig:clip_comp}
\end{figure}

We designed our classifier $F_{\Theta}$ using a deep fully-connected neural network without convolutional layers (\ie~a MLP).
As illustrated in \mbox{\Cref{fig:clip_comp}(c)}, $F_{\Theta}$ takes three different inputs: the input image embedding, the input sentence embedding, and the discrete demographics vector (\textit{age: three levels} and \textit{occupation: three categories}) and the output are five usability ratings of semantic distance and familiarity.
We obtained the image and sentence embedding using the image encoder and the text decoder of IconCLIP.
And $F_{\Theta}$ process these inputs with four fully-connected layers (using ReLU activations and 256
channels per layer.

\subsubsection{Visual Distinguishability}
As discussed in \Cref{sec:interface_viz}, \name~provides a distinguishability graph to help users compare the relative visual distance between icons in the prepared icon set.
We directly use the embedding space of the finetuned IconCLIP as our similarity measurement space.
We used Uniform Manifold Approximation and Projection (UMAP)~\cite{mcinnes2018umap-software} to project the $512$d feature vector to $2$d.
\begin{figure}[t!]
\centering
  \includegraphics[width=\linewidth]{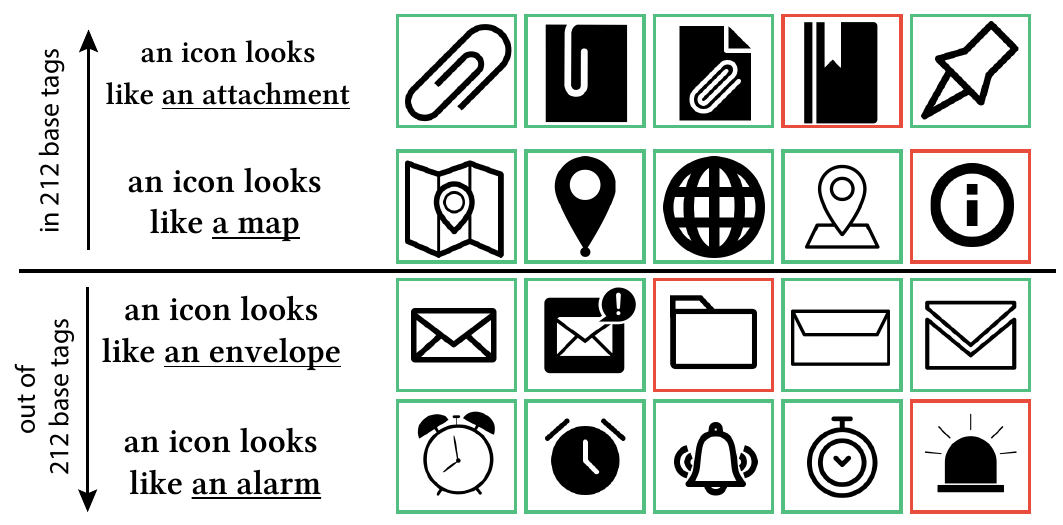}
  \caption{
      \textbf{text-to-image retrieval by IconCLIP.}
      Among the four tags we used as queries, only ``attachment'' and ``map'' are in the $212$ base tags.
      The IconCLIP embedding space recognizes the meaning of ``envelop'' and ``alarm'' because we used the tags associated with icons (we use the green box for positive results and the red box for negative results).
    }
\label{fig:vis_re_eval}
\end{figure}

\section{Evaluation}
\subsection{Evaluation of IconCLIP}
\label{sec:iconclip_eval}
To evaluate the IconCLIP embedding space, we performed a top-k image retrieval evaluation.
We split the overall collected icons into a training set and a testing set.
We used the training set to finetune IconCLIP, and we performed the retrieval test as follows: we used each icon image in the testing set as a query and used the rest of the testing set as a retrieval set.
And we consider a retrieved icon image as a positive result if it shares a common tag with the query image.
The MAP$@$5 (mean average precision at rank $5$) of the retrieval test is $74.3$.
On the other hand, we also performed a text-to-image retrieval test and showed the qualitative image retrieval results in \mbox{\Cref{fig:vis_re_eval}}.
We can observe that the top-5 nearest neighbors match the tags in the sentence even when the concepts are not in the $212$ base tags we used for collecting the icons.
This suggests that the tags associated with the icons expand the embedding space of IconCLIP.
\subsection{Evaluation of perceptual usability feedback models}
As mentioned in \Cref{sec:perp_predict}, we trained a unified network to predict usability ratings based on the icon's image embeddings, tag embeddings, and demographic information.
To demonstrate the ability to predict usability ratings of unseen tags, we split the icons of $50$ base tags into $45$ tags as seen and $5$ tags as unseen tags.
It should be noted that we only use the base tags as the selection criteria, but we used all the associated tags within the base tags as training signals, so it is not restricted to these base tags. 
We performed two types of evaluation of our prediction model.
\subsubsection{In-domain Evaluation}
First, we evaluated the prediction precision and recall on the $45$ base functions we used for training.
Among the icons of these $45$ base functions, we randomly split the data into $90/10$ as training/testing data.
For \textit{semantic distance}, our models achieved $83.6\%$ for precision and $84.1\%$ for recall.
For \textit{familiarity}, our models achieved $76.3\%$ for precision and $77.6\%$ for recall.
\subsubsection{Out-of-domain Evaluation}
Second, we also evaluated the prediction precision and recall on all icons belonging to the $5$ base tags we held out during training.
For \textit{semantic distance}, our models achieved $66.4\%$ for precision and $69.5\%$ for recall.
For \textit{familiarity}, our models achieved $67.1\%$ for precision and $68.4\%$ for recall.

\begin{figure*}[h!]
    \centering
    \includegraphics[width=\linewidth]{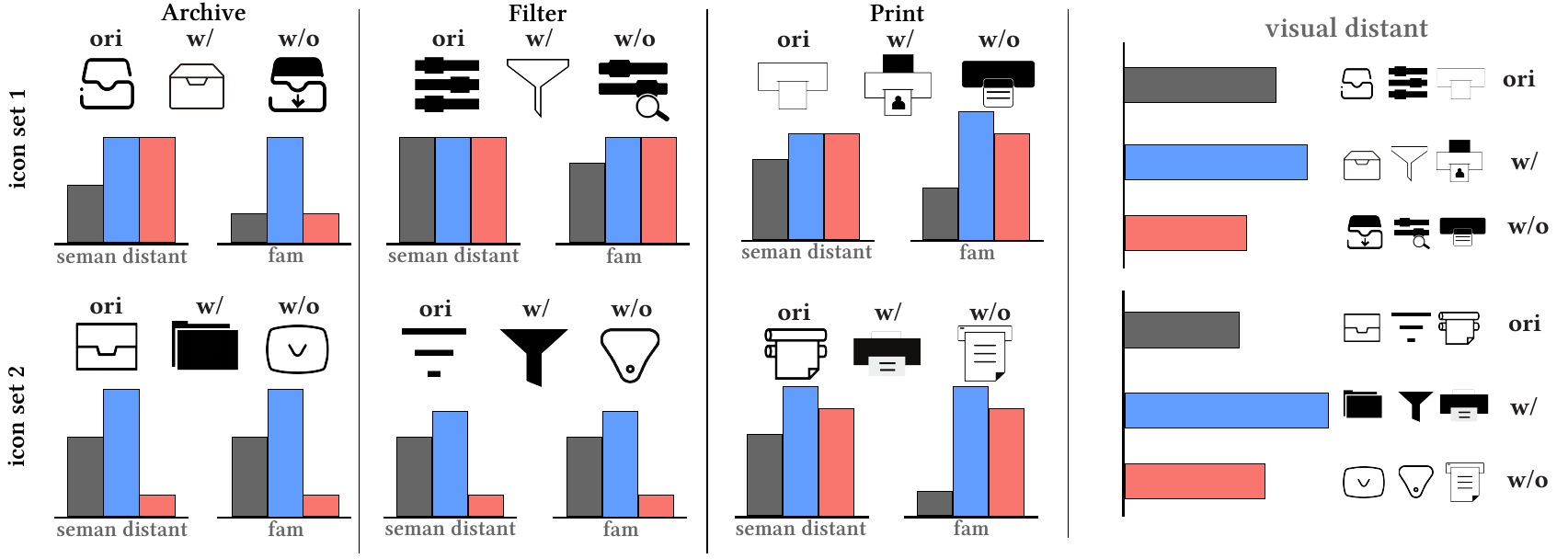}
    \caption{
    The six original icons and their examples of revised icons with and without \name. 
    We plot the crowdsourced evaluation results (``AMT rating'') of each icon.
    The \textcolor{evalgray}{gray}/\textcolor{evalblue}{blue}/\textcolor{evalred}{red} bar denotes the AMT rating of \textcolor{evalgray}{original icon}/\textcolor{evalblue}{icon revised with \name}/\textcolor{evalred}{icon revised without \name}.
    The ratings ranged from 1 (``Very Bad'') to 5 (``Very Good'') for the bar chart of semantic distance (seman distant) and familiarity (fam).
    We can see that most of the icons revised with \name~ received higher AMT ratings than icons revised without \name. 
    We also show the visual distinguishability score between each icon in the embedding space.
    The visual distinguishability between icons revised with EvIcon is the furthest.
    }
    \label{fig:revision_evaluation}
\end{figure*}
\subsection{Evaluation with UI Designers}
\label{sec:eva}
To evaluate how EvIcon's interaction design can support designers' revision process, we conducted a user study with six professional UI designers (five females and one male; ages ranging from 22 to 34 years old).
We recruited a similar number of domain experts with prior similar works~\cite{Peng:2018:AS,Xing2016,SimoSerraSIGGRAPH2018}. 
The self-reported professional experience of the designers ranges from one to ten years.
All of them used Adobe Illustrator\footnote{\url{https://www.adobe.com/products/illustrator.html}} for initial icon design.
\subsubsection{Procedure and Tasks}
\label{subsec:procedure}
After introducing \name~and the meaning of two types of feedback, the designers practiced using \name~for ten minutes. 
We then asked them to complete the practice tasks (\eg~reporting the perceptual usability of an icon in different age groups of users) to ensure they understand how to use \name. 
In the formal sessions, the designers were asked to improve the usability of two icon sets. For the design brief to guide the designers when revising icons, we informed the designers the scenario of the evaluation is that a client asked them to improve the icon sets so that these icons can be used in a wide range of applications and users (e.g., elders).
Each icon set contains three icons of tags ``Archive'', ``Print'', and ``Filter''.
We selected these tags based on their average familiarity level collected via the crowdsourced study in \Cref{sec:dataset} (``Archive'': 3.8; ``Filter'': 3.9; ``Print'': 4.2) to ensure we included established and uncommon tags in the evaluation. 
Moreover, ``Archive'' and ``Print'' are in the $45$ seen tag set, and ``Filter'' is in the $5$ unseen tag set used in \Cref{sec:iconclip_eval}.
We denoted these icons as the original icons in the following sections.

As shown in \Cref{fig:revision_evaluation}, the icons in the two sets are different, and we instructed each designer to improve the usability of one icon set with \name~and another set without \name, both in fifteen minutes. The combination of the icon set and two conditions were randomly assigned, and the order of conditions was counterbalanced to avoid the learning effect.
Under both conditions, the designers can freely edit icons using the design tool of their choice and search online for the information. 
However, under the without \mbox{\name}~condition, the designers can not access the icons and perceptual ratings we collected.
We recorded the revision process and the revised icons. 
In the end, we obtained 36 revised icons from six designers in total, and we found that all designers spent the entire time budget (15 minutes) for each condition.

\begin{figure*}[h!]
    \centering
    \includegraphics[width=\linewidth]{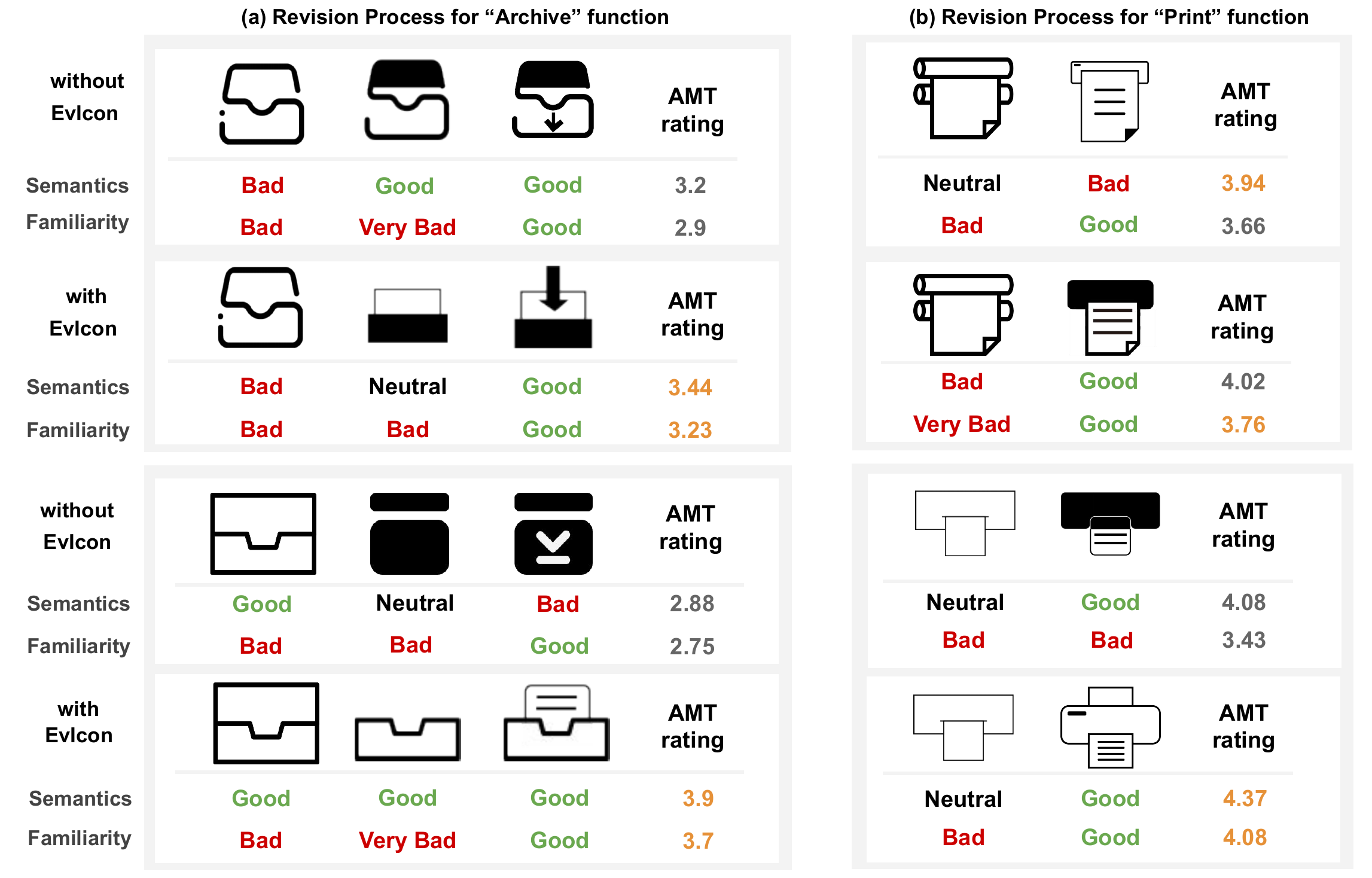}
    \caption{
    Icon revision process starting from left to right by designers with and without \name. (a) presents the revision processes for ``Archive'' icons of two designers, and (b) presents the revision processes for ``Print'' icons of two designers. 
    Eight groups of revision process with the prediction of perception feedback and the crowdsourced evaluation results (``AMT rating'') of the finalized icons are presented.  
    }
    \label{fig:revision_sequence}
\end{figure*}
\subsubsection{Result}
\begin{description}[style=unboxed,leftmargin=0cm]
\item[Revised icons]
In \Cref{fig:revision_evaluation}, we show the revised icons of all three tags with and without using \name.
To further verify that \name~ can help designers improve icons' usability, we launched an additional crowdsourced evaluation on AMT to collect $213$ ($140$ males and $73$ females; $19$ to $64$ years old) crowdworkers' usability ratings of all original and revised icons pair in an assignment.
We collected averaged $57.8$ unique ratings for each original/revised icon pair. 
Each crowdworker would only rate an icon pair revised by the same designers to eliminate the influence of individual designers' abilities. 
We used the majority vote of all received ratings as the final rating of each revised icon to reduce the effects of spammers as shown in \Cref{fig:revision_evaluation}.
To reduce the mutual influence of icons in different pairs of revised and original icons, we calculated the final ratings of the original icons by averaging the majority vote ratings across the different pairs of revised icons provided by the designers.

We can see from \Cref{fig:revision_evaluation} that most of the revised icons with \name~ (blue bars) obtained higher AMT ratings of semantic distance and familiarity than those without \name.
Moreover, to demonstrate the usefulness of \name~in improving the visual distinguishability within the icon set, we computed the mutual distances between the 512-dim embedded vector of each revised icon.
In the rightmost panel of \Cref{fig:revision_evaluation}, the mutual distance between icons revised with \name~is farther than the original icons and icons revised without \name, which suggests better visual distinguishability.

\Cref{fig:revision_sequence} illustrates example revision processes for the icons ``Archive'' and ``Print.'' Throughout each design step, \name~provided feedback on ``Semantics'' (semantic distance) and ``Familiarity.'' The crowdsourced evaluation results, collected via Amazon Mechanical Turk (AMT), are displayed next to the finalized icon (the right-most icon of each block in \Cref{fig:revision_sequence}) for each revision process. The evaluation outcomes demonstrate that icons revised with \name~generally outperformed those revised without it, as shown in \Cref{fig:revision_sequence}, with higher ratings given for both ``Semantics'' and ``Familiarity.''

\item[Revised icons for diverse demographics]
We investigated whether icons revised with \name result in higher usability ratings from older users ($>50$ years old) to demonstrate the tool's ability to create more inclusive designs for users with diverse demographics
We found that the revised icons obtained higher AMT mean semantic distance (with: $3.49$; without: $3.35$) and familiarity (with: $2.83$; without: $2.81$) ratings across tags.
We show the two examples of the revised icons using \name~in \Cref{fig:elder_print}. 

\begin{figure}[h!]
\centering
  \includegraphics[width=\linewidth]{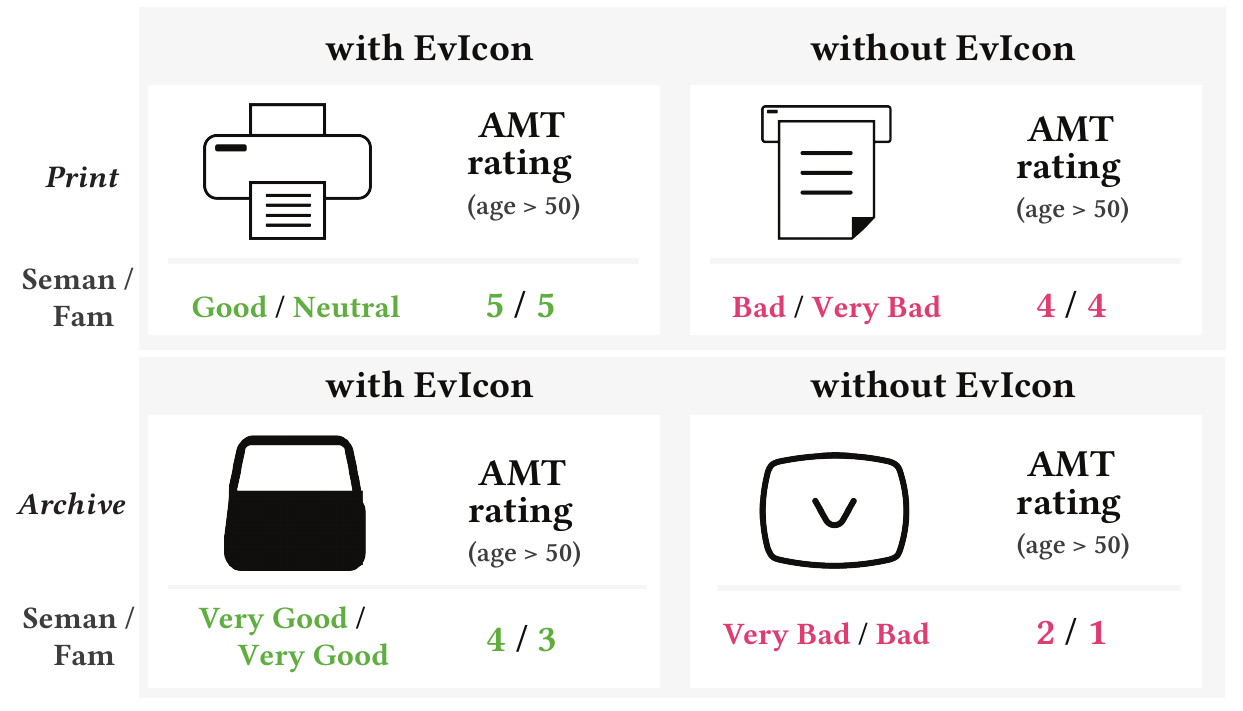}
  \caption{
  We show two examples of revised icons with and without \name. We can see that the icons revised by \name with better predicted semantic distance and familiarity levels also achieved higher AMT ratings from the elder crowdworkers (age $>$ 50).}
\label{fig:elder_print}
\end{figure}

\item[Crowdsourced evaluation on revised icons]
\begin{figure}
\centering
  \includegraphics[width=1\linewidth]{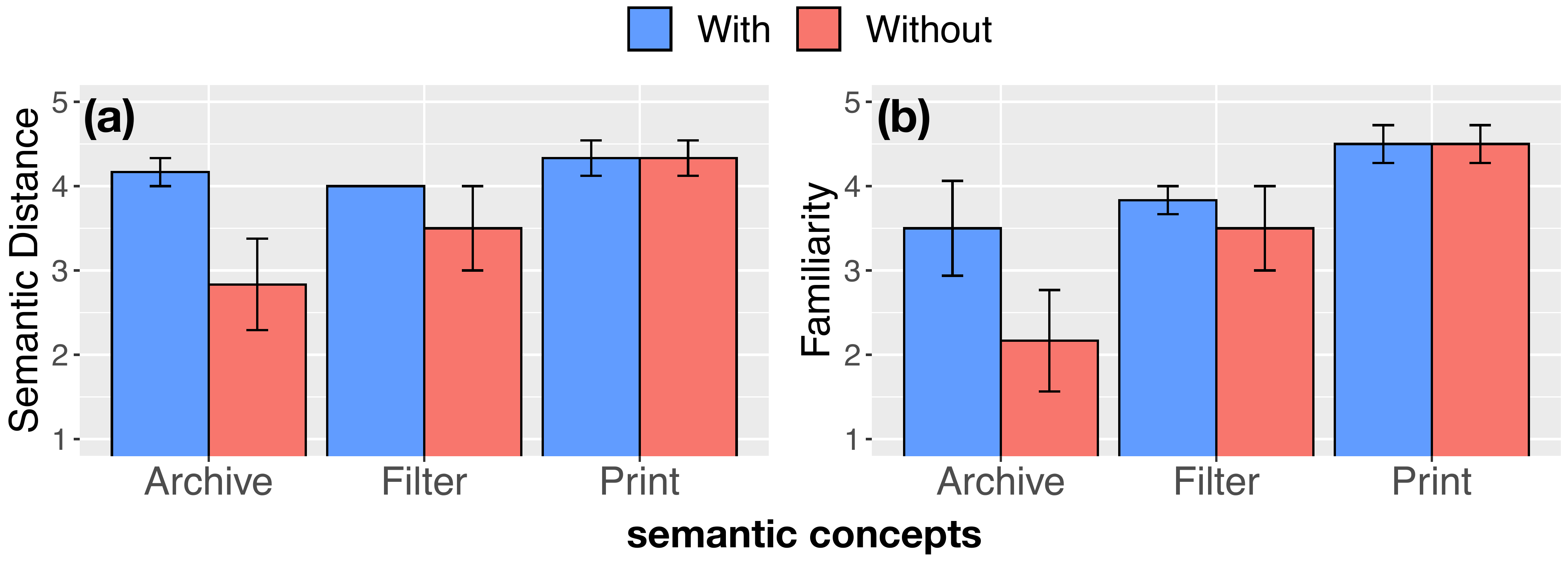}
  \caption{
  The AMT ratings of the icons revised by all designers. 
  (a) The ratings of semantic distance rating. (b) The ratings of familiarity rating. 
  The error bars represent the standard deviation.}
\label{fig:amt_eva}
\end{figure}
As shown in \mbox{\Cref{fig:amt_eva}}, we compared the averaged mode ratings of the revised icons by their tag and whether they were revised with \mbox{\name} using Cohen's \textit{d}. 
We can see that for the icons revised by all designers, the ``Archive'' and ``Filter'' icons revised with \mbox{\name}~ received a higher level of semantic distance (Archive: \textit{d} = 1.48; Filter: \textit{d} = 0.63; \mbox{\Cref{fig:amt_eva}(a)}) and familiarity (Archive: \textit{d} = 1.02; Filter: \textit{d} = 0.4; \mbox{\Cref{fig:amt_eva}(b)}) than those without using \mbox{\name} with the moderate to the large magnitude of the mean difference.
Yet, the ``Print'' icons revised with \mbox{\name}~ obtained the same level of semantic distance and familiarity as those without \mbox{\name}.
These results suggest that the designers benefit most from using \mbox{\name}~ in improving the usability of icons of unestablished tags (\ie~``Archive'' and ``Filter'').
Since most of the designers and users have not formed the common visual metaphors for the unestablished tags, \mbox{\name's} feedback helps designers navigate the vast variations of ``Archive'' and ``Filter'' icons and find the best way to revise the icons.

\item[Revised icon retrieval test]
The retrieval test aims to demonstrate that the existing icons in our dataset are used primarily as inspiration rather than copied directly. In \Cref{fig:retrieval}, we present the closest example from our dataset for each revised icon and provide the PSNR/SSIM scores. We observe that for the semantic concept with simpler shapes, such as ``Filter,'' the revised icons are generally closer to the existing icons in our dataset. However, for the semantic concept with more complex shapes, designers tend to make more significant revisions (e.g., ``Print'', ``Archive''), resulting in greater distances between the revised icons and their closest examples.

\begin{figure*}[h!]
\centering
  \includegraphics[width=\linewidth]{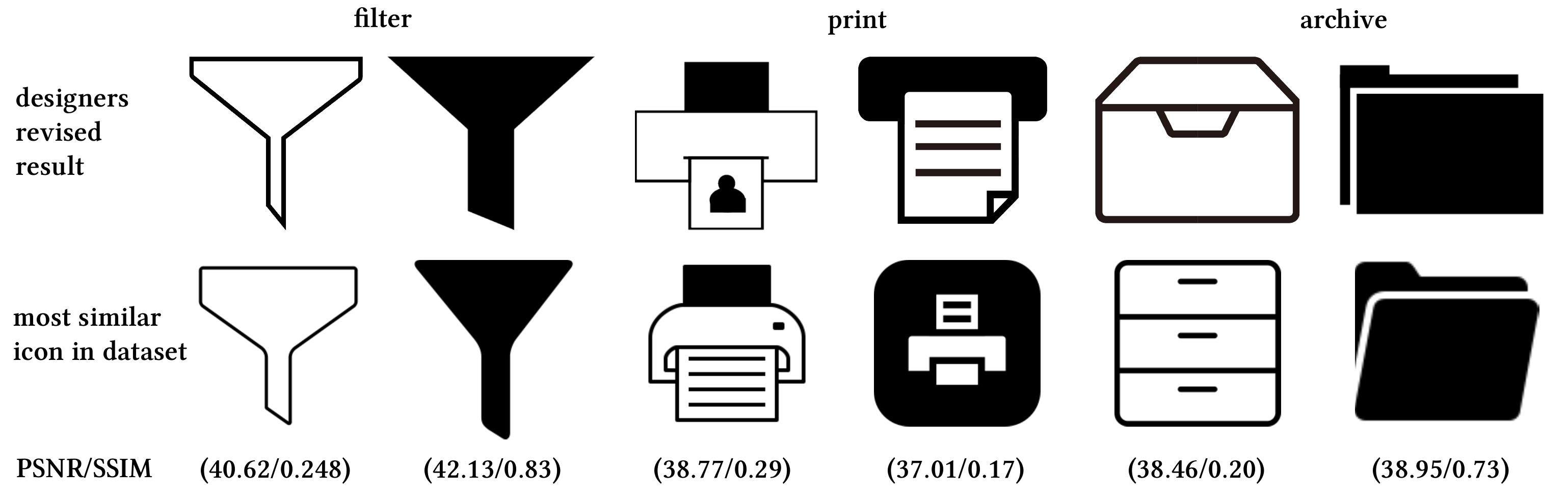}
  \caption{
  For each icon revised by the designers, we show its closest example in our dataset and its corresponding PSNR/SSIM at the bottom.
  }
\label{fig:retrieval}
\end{figure*}

\item[Post-study interview]
In the post-study interviews, all six designers gave positive attitudes towards \name. 
The designers mentioned that when revising icons with \name, they got the idea of how to revise an icon to meet public understanding more easily by checking the perception feedback constantly.
They found the perception feedback convincing as it was generated based on data labeled by over two thousand crowdworkers:
\begin{itemize}
    \item \textit{``\name~ keeps me on the right track and ensures that my design can be understood by others while I modify the icon design based on my creativity. 
    ''}(\textit{P3})
    \item \textit{``The good or bad rating provided by the system is promising and helpful in designing high-usability interface icons, compared to designing the icons on my own.''}(\textit{P5})
\end{itemize}
Some designers were amazed by the perception feedback for specific demographics since they have experienced struggling to design interface icons targeting a specific category of users while having limited knowledge or access to the users:
\begin{itemize}
\item \textit{``The feedback from a specific demographic is very useful. 
I can adjust the icons according to the feedback from my target user's category provided by the system.
This tool definitely helps this.''}(\textit{P4})
\item \textit{``I am touched to see how this tool supports elders' feedback! 
Although icons play an important role in interface design, there is not much information about which icons are friendly or recognizable to elders. 
''}(\textit{P6})
\end{itemize}
Designers also found the distinguishability visualization panel helpful. Both P2 and P6 said they would check the related distance between the uploaded icon and the icons in the suggestion panel to see how they could improve their design.
P2, P3, P5, and P6 mentioned they could derive some graphical design features from the icon suggestion panel that can be added to their own designs:
\begin{itemize}
\item \textit{``It is interesting that the system provides designs from other designers based on current target function.''}(\textit{P3})
\item \textit{``I can see those good icons in the suggestion panel, and think about how to start my design based on the recommendations. 
It will help save my time to grasp users' thoughts at the beginning of the design flow.''}(\textit{P5})
\end{itemize}

Designers also discussed possible benefits \name~ could bring if applied in their current workflow. 
P5 said it would save lots of time to notice the perception gap between designers, engineers, and average users earlier with \name, instead of finding out in usability testing after several design iterations and discussions. 
As designers, participants usually care a lot about aesthetics while designing icons, \name~could also provide assistances to balance between aesthetics and usability.
\begin{itemize}
\item \textit{``It was nice that I could see the perception differences between public users and my personal thoughts and styles.''}(\textit{P2})
\item \textit{``Designers often want to design an aesthetic and unique icon, but sometimes they go too far that the icon becomes unrecognizable to users. 
With \name, it would be easier to take both aesthetic and usability into consideration at the same time.''}(\textit{P3})
\item \textit{``Designers often add more styling details in the later phase of the iteration and worsen the icons' distinguishability. 
With \name~, we can check the perception feedback in each iteration to ensure the quality of our designed icons.''}(\textit{P4})
\end{itemize}
The designers also mentioned that the perception feedback could improve communication with their colleagues or clients if \name~ is included in their design process.
\begin{itemize}
\item \textit{``I could convince the clients that my design is good with \name.''}(\textit{P3})
\item \textit{``The results from \name~ would be a promising report to defend our design against clients.''}(\textit{P4})
\end{itemize}
The designers confirmed that \name~ could generally be useful and mentioned possibilities of how \name~ can assist in different design phases.
Moreover, they are willing to use \name~ in their design process if it becomes a mature product in the future.

\end{description}
\section{Limitations and Future Work}
\begin{description}[style=unboxed,leftmargin=0cm]
\item[Single icon style]
As flat design continues to be a popular trend in digital design, our framework currently focuses on improving the usability of single-colored icons. However, we recognize the importance of expanding EvIcon's capabilities to include a wider range of icon styles. To achieve this, we plan to build a diverse dataset of icons and use metrics proposed in \cite{icon_style} to explore and compare icons in different styles. By doing so, we aim to improve the generalizability of EvIcon and make it more adaptable to the changing trends and preferences in digital design. Expanding the dataset and incorporating new metrics will enable EvIcon to produce icons in a wider variety of styles, ensuring that it remains a valuable tool for UI designers across different industries and contexts.

\item[User interface with limited icon editing functions]
While EvIcon's vector graphics editor offers a basic set of design tools, it may not be sufficient for the needs of some UI designers. The absence of advanced features could limit creativity and lead to a less efficient workflow. Our plan to integrate EvIcon as a plugin for professional design tools such as Adobe Illustrator and Sketch\footnote{\url{https://www.sketch.com/}} is aimed at addressing these limitations. By providing access to a more comprehensive suite of design tools, UI designers can expand their creative options and improve their efficiency. The integration will enable designers to access EvIcon's icon creation and editing features within their preferred design software, eliminating the need to switch between multiple tools. Ultimately, this will enable designers to produce higher quality icons more efficiently, resulting in better user experiences for their products.

\item[Supporting validations for general use of icons]
While the proposed framework is primarily focused on supporting designers in validating and revising icons for user interface design, it is important to note that icons have many other applications beyond UI design. Icons are widely used in presentation slides, infographics, and other forms of visual communication, where their design requirements may differ from those in UI design. For example, the icons used in infographics may require better abilities to convey information rather than better familiarity with viewers.
Given the versatile nature of icons, we plan to extend the usage scenario and target audience of EvIcon to support icon improvement and selection for more general purposes. By doing so, we aim to make EvIcon a more versatile tool that can assist designers across various fields and contexts, not just limited to UI design. Expanding EvIcon's capabilities to accommodate different design requirements and user needs will enhance its value and relevance, making it an even more valuable tool for designers working on a wide range of projects. Ultimately, this will enable designers to create more effective and engaging visual content across various domains, resulting in better user experiences for their audiences.

\end{description}
\section{Conclusion}

In this paper, we propose a human-in-the-loop framework called \name that aims to enhance the usability of interface icon sets. Our framework includes a novel perceptual usability formulation and an interactive design tool that enable users to modify icons' effectiveness in conveying information. We also introduce the first icon dataset, \textit{IconCEPT10K}, which features high-level perceptual usability ratings, such as semantic distance and familiarity, from over 5,000 crowdworkers. To demonstrate the effectiveness of \name, we conducted a user study with six UI designers. Our quantitative and qualitative results show that using \name resulted in an icon set with improved usability, as rated by over 200 crowdworkers. These findings suggest that \name is an effective tool for facilitating the design process.

\bibliographystyle{eg-alpha-doi} 
\bibliography{paper}       

\newcommand{\etalchar}[1]{$^{#1}$}
\begin{thebibliography}{\uppercase{SABAG{\etalchar{*}}05}}

\bibitem[AMW21]{ali2021anachronism}
\textsc{Ali A.~X., Mcaweeney E., Wobbrock J.~O.}:
\newblock Anachronism by design: Understanding young adults’ perceptions of
  computer iconography.
\newblock \emph{International Journal of Human-Computer Studies} (2021),
  102599.

\bibitem[Arl14]{arledge2014filled}
\textsc{Arledge C.}:
\newblock Filled-in vs. outline icons: the impact of icon style on usability.

\bibitem[AV07]{kmeans}
\textsc{Arthur D., Vassilvitskii S.}:
\newblock K-means++: The advantages of careful seeding.
\newblock In \emph{Proceedings of the Eighteenth Annual ACM-SIAM Symposium on
  Discrete Algorithms} (USA, 2007), SODA '07, Society for Industrial and
  Applied Mathematics, p.~1027–1035.

\bibitem[AZF{\etalchar{*}}21]{abdal2021clip2stylegan}
\textsc{Abdal R., Zhu P., Femiani J., Mitra N.~J., Wonka P.}:
\newblock Clip2stylegan: Unsupervised extraction of stylegan edit directions.
\newblock \emph{arXiv preprint arXiv:2112.05219} (2021).

\bibitem[BAR92]{bailey1992usability}
\textsc{Bailey R.~W., Allan R.~W., Raiello P.}:
\newblock Usability testing vs. heuristic evaluation: A head-to-head
  comparison.
\newblock In \emph{Proceedings of the human factors society annual meeting}
  (1992), vol.~36, SAGE Publications Sage CA: Los Angeles, CA, pp.~409--413.

\bibitem[BBDF10]{brochu2010bayesian}
\textsc{Brochu E., Brochu T., De~Freitas N.}:
\newblock A bayesian interactive optimization approach to procedural animation
  design.
\newblock In \emph{Proceedings of the 2010 ACM SIGGRAPH/Eurographics Symposium
  on Computer Animation} (2010), pp.~103--112.

\bibitem[BKO{\etalchar{*}}17]{bylinskii2017learning}
\textsc{Bylinskii Z., Kim N.~W., O'Donovan P., Alsheikh S., Madan S., Pfister
  H., Durand F., Russell B., Hertzmann A.}:
\newblock Learning visual importance for graphic designs and data
  visualizations.
\newblock In \emph{Proceedings of the 30th Annual ACM symposium on user
  interface software and technology} (2017), pp.~57--69.

\bibitem[BL15]{BL:2015:lillicon}
\textsc{Bernstein G.~L., Li W.}:
\newblock Lillicon: Using transient widgets to create scale variations of
  icons.
\newblock \emph{ACM Transactions on Graphics (TOG) 34}, 4 (2015), 1--11.

\bibitem[CGW{\etalchar{*}}14]{Chen:2014:HAV}
\textsc{Chen H.-T., Grossman T., Wei L.-Y., Schmidt R.~M., Hartmann B.,
  Fitzmaurice G., Agrawala M.}:
\newblock History assisted view authoring for 3d models.
\newblock In \emph{Proceedings of the SIGCHI Conference on Human Factors in
  Computing Systems} (New York, NY, USA, 2014), CHI '14, ACM, pp.~2027--2036.
\newblock URL: \url{http://doi.acm.org/10.1145/2556288.2557009}, \href
  {https://doi.org/10.1145/2556288.2557009}
  {\path{doi:10.1145/2556288.2557009}}.

\bibitem[CLKL16]{cherng2016eeg}
\textsc{Cherng F.-Y., Lin W.-C., King J.-T., Lee Y.-C.}:
\newblock An eeg-based approach for evaluating graphic icons from the
  perspective of semantic distance.
\newblock In \emph{Proceedings of the 2016 chi conference on human factors in
  computing systems} (2016), ACM, pp.~4378--4389.

\bibitem[CSSI21]{chong2021interactive}
\textsc{Chong T., Shen I.-C., Sato I., Igarashi T.}:
\newblock Interactive optimization of generative image modelling using
  sequential subspace search and content-based guidance.
\newblock In \emph{Computer Graphics Forum} (2021), vol.~40, Wiley Online
  Library, pp.~279--292.

\bibitem[CUG20]{chajadi2020effects}
\textsc{Chajadi F., Uddin M.~S., Gutwin C.}:
\newblock Effects of visual distinctiveness on learning and retrieval in icon
  toolbars.

\bibitem[DBK{\etalchar{*}}21]{dosovitskiy2020vit}
\textsc{Dosovitskiy A., Beyer L., Kolesnikov A., Weissenborn D., Zhai X.,
  Unterthiner T., Dehghani M., Minderer M., Heigold G., Gelly S., Uszkoreit J.,
  Houlsby N.}:
\newblock An image is worth 16x16 words: Transformers for image recognition at
  scale.
\newblock \emph{ICLR} (2021).

\bibitem[DHF{\etalchar{*}}17a]{Rico}
\textsc{Deka B., Huang Z., Franzen C., Hibschman J., Afergan D., Li Y., Nichols
  J., Kumar R.}:
\newblock Rico: A mobile app dataset for building data-driven design
  applications.
\newblock UIST '17, Association for Computing Machinery, p.~845–854.
\newblock URL: \url{https://doi.org/10.1145/3126594.3126651}, \href
  {https://doi.org/10.1145/3126594.3126651}
  {\path{doi:10.1145/3126594.3126651}}.

\bibitem[DHF{\etalchar{*}}17b]{deka2017zipt}
\textsc{Deka B., Huang Z., Franzen C., Nichols J., Li Y., Kumar R.}:
\newblock Zipt: Zero-integration performance testing of mobile app designs.
\newblock In \emph{Proceedings of the 30th Annual ACM Symposium on User
  Interface Software and Technology} (2017), pp.~727--736.

\bibitem[DHF{\etalchar{*}}17c]{ZIPT}
\textsc{Deka B., Huang Z., Franzen C., Nichols J., Li Y., Kumar R.}:
\newblock Zipt: Zero-integration performance testing of mobile app designs.
\newblock In \emph{Proceedings of the 30th Annual ACM Symposium on User
  Interface Software and Technology} (New York, NY, USA, 2017), UIST '17,
  Association for Computing Machinery, p.~727–736.
\newblock URL: \url{https://doi.org/10.1145/3126594.3126647}, \href
  {https://doi.org/10.1145/3126594.3126647}
  {\path{doi:10.1145/3126594.3126647}}.

\bibitem[GAHG17]{icon_style}
\textsc{Garces E., Agarwala A., Hertzmann A., Gutierrez D.}:
\newblock Style-based exploration of illustration datasets.
\newblock \emph{Multimedia Tools Appl. 76}, 11 (jun 2017), 13067–13086.

\bibitem[Git86]{gittins1986icon}
\textsc{Gittins D.}:
\newblock Icon-based human-computer interaction.
\newblock \emph{International Journal of Man-Machine Studies 24}, 6 (1986),
  519--543.

\bibitem[GSF01]{goonetilleke2001effects}
\textsc{Goonetilleke R.~S., Shih H.~M., FRITSCH J.}:
\newblock Effects of training and representational characteristics in icon
  design.
\newblock \emph{International Journal of Human-Computer Studies 55}, 5 (2001),
  741--760.

\bibitem[Hol16]{holzinger2016interactive}
\textsc{Holzinger A.}:
\newblock Interactive machine learning for health informatics: when do we need
  the human-in-the-loop?
\newblock \emph{Brain Informatics 3}, 2 (2016), 119--131.

\bibitem[Hor94]{Horton1994}
\textsc{Horton W.~K.}:
\newblock \emph{The ICON Book: Visual Symbols for Computer Systems and
  Documentation}.
\newblock John Wiley \& Sons, Inc., USA, 1994.

\bibitem[Hor96]{Horton1996}
\textsc{Horton W.}:
\newblock Designing icons and visual symbols.
\newblock In \emph{Conference Companion on Human Factors in Computing Systems}
  (New York, NY, USA, 1996), CHI '96, Association for Computing Machinery,
  p.~371–372.
\newblock URL: \url{https://doi.org/10.1145/257089.257378}, \href
  {https://doi.org/10.1145/257089.257378} {\path{doi:10.1145/257089.257378}}.

\bibitem[Hsi17]{hsieh2017multiple}
\textsc{Hsieh T.-J.}:
\newblock Multiple roles of color information in the perception of icon-type
  images.
\newblock \emph{Color Research \& Application 42}, 6 (2017), 740--752.

\bibitem[IMC07]{isherwood2007icon}
\textsc{Isherwood S.~J., McDougall S.~J., Curry M.~B.}:
\newblock Icon identification in context: The changing role of icon
  characteristics with user experience.
\newblock \emph{Human Factors: The Journal of the Human Factors and Ergonomics
  Society 49}, 3 (2007), 465--476.

\bibitem[KFZI20]{kamarulzaman2020comparative}
\textsc{Kamarulzaman N.~A., Fabil N., Zaki Z.~M., Ismail R.}:
\newblock Comparative study of icon design for mobile application.
\newblock In \emph{Journal of Physics: Conference Series} (2020), vol.~1551,
  IOP Publishing, p.~012007.

\bibitem[KPL08]{KOLHOFF2008550}
\textsc{Kolhoff P., Preuß J., Loviscach J.}:
\newblock Content-based icons for music files.
\newblock \emph{Computers \& Graphics 32}, 5 (2008), 550--560.
\newblock URL:
  \url{https://www.sciencedirect.com/science/article/pii/S009784930800006X},
  \href {https://doi.org/https://doi.org/10.1016/j.cag.2008.01.006}
  {\path{doi:https://doi.org/10.1016/j.cag.2008.01.006}}.

\bibitem[KRG13]{komarov2013crowdsourcing}
\textsc{Komarov S., Reinecke K., Gajos K.~Z.}:
\newblock Crowdsourcing performance evaluations of user interfaces.
\newblock In \emph{Proceedings of the SIGCHI conference on human factors in
  computing systems} (2013), pp.~207--216.

\bibitem[KS96]{ketchen1996application}
\textsc{Ketchen D.~J., Shook C.~L.}:
\newblock The application of cluster analysis in strategic management research:
  an analysis and critique.
\newblock \emph{Strategic management journal 17}, 6 (1996), 441--458.

\bibitem[KSG20]{koyama2020sequential}
\textsc{Koyama Y., Sato I., Goto M.}:
\newblock Sequential gallery for interactive visual design optimization.
\newblock \emph{ACM Transactions on Graphics (TOG) 39}, 4 (2020), 88--1.

\bibitem[KSI14]{koyama2014crowd}
\textsc{Koyama Y., Sakamoto D., Igarashi T.}:
\newblock Crowd-powered parameter analysis for visual design exploration.
\newblock In \emph{Proceedings of the 27th annual ACM symposium on User
  interface software and technology} (2014), pp.~65--74.

\bibitem[KSSI17]{yuki2017sls}
\textsc{Koyama Y., Sato I., Sakamoto D., Igarashi T.}:
\newblock Sequential line search for efficient visual design optimization by
  crowds.
\newblock URL: \url{https://doi.org/10.1145/3072959.3073598}, \href
  {https://doi.org/10.1145/3072959.3073598}
  {\path{doi:10.1145/3072959.3073598}}.

\bibitem[Kur00]{kurniawan2000rule}
\textsc{Kurniawan S.~H.}:
\newblock A rule of thumb of icons' visual distinctiveness.
\newblock In \emph{Proceedings on the 2000 conference on Universal Usability}
  (2000), pp.~159--160.

\bibitem[LC20]{legleiter2020flat}
\textsc{Legleiter A.~M., Caporusso N.}:
\newblock Flat-design icon sets: A case for universal meanings?
\newblock In \emph{International Conference on Applied Human Factors and
  Ergonomics} (2020), Springer, pp.~211--217.

\bibitem[LCS{\etalchar{*}}18]{liu2018learning}
\textsc{Liu T.~F., Craft M., Situ J., Yumer E., Mech R., Kumar R.}:
\newblock Learning design semantics for mobile apps.
\newblock In \emph{The 31st Annual ACM Symposium on User Interface Software and
  Technology} (2018), ACM, pp.~569--579.

\bibitem[LGG18]{lagunas2018learning}
\textsc{Lagunas M., Garces E., Gutierrez D.}:
\newblock Learning icons appearance similarity.
\newblock \emph{Multimedia Tools and Applications} (2018), 1--19.

\bibitem[LGM20]{liao2020questioning}
\textsc{Liao Q.~V., Gruen D., Miller S.}:
\newblock Questioning the ai: informing design practices for explainable ai
  user experiences.
\newblock In \emph{Proceedings of the 2020 CHI Conference on Human Factors in
  Computing Systems} (2020), pp.~1--15.

\bibitem[Lin94]{lin1994study}
\textsc{Lin R.}:
\newblock A study of visual features for icon design.
\newblock \emph{Design Studies 15}, 2 (1994), 185--197.
\newblock URL:
  \url{https://www.sciencedirect.com/science/article/pii/0142694X94900248},
  \href {https://doi.org/https://doi.org/10.1016/0142-694X(94)90024-8}
  {\path{doi:https://doi.org/10.1016/0142-694X(94)90024-8}}.

\bibitem[LKC{\etalchar{*}}16]{laursen2016icon}
\textsc{Laursen L.~F., Koyama Y., Chen H.-T., Garces E., Gutierrez D., Harper
  R., Igarashi T.}:
\newblock Icon set selection via human computation.

\bibitem[LMG11]{leung2011age}
\textsc{Leung R., McGrenere J., Graf P.}:
\newblock Age-related differences in the initial usability of mobile device
  icons.
\newblock \emph{Behaviour \& Information Technology 30}, 5 (2011), 629--642.

\bibitem[LRFN04]{Lewis2004icons}
\textsc{Lewis J.~P., Rosenholtz R., Fong N., Neumann U.}:
\newblock Visualids: Automatic distinctive icons for desktop interfaces.
\newblock \emph{ACM Trans. Graph. 23}, 3 (Aug. 2004), 416–423.
\newblock URL: \url{https://doi.org/10.1145/1015706.1015739}, \href
  {https://doi.org/10.1145/1015706.1015739}
  {\path{doi:10.1145/1015706.1015739}}.

\bibitem[LZC11]{Lee:2011:SRU}
\textsc{Lee Y.~J., Zitnick C.~L., Cohen M.~F.}:
\newblock Shadowdraw: Real-time user guidance for freehand drawing.
\newblock \emph{ACM Trans. Graph. 30}, 4 (July 2011), 27:1--27:10.
\newblock URL: \url{http://doi.acm.org/10.1145/2010324.1964922}, \href
  {https://doi.org/10.1145/2010324.1964922}
  {\path{doi:10.1145/2010324.1964922}}.

\bibitem[MCdB99]{mcdougall1999measuring}
\textsc{Mcdougall S.~J., Curry M.~B., de~Bruijn O.}:
\newblock Measuring symbol and icon characteristics: Norms for concreteness,
  complexity, meaningfulness, familiarity, and semantic distance for 239
  symbols.
\newblock \emph{Behavior Research Methods, Instruments, \& Computers 31}, 3
  (1999), 487--519.

\bibitem[MCdB01]{mcdougall2001effects}
\textsc{McDougall S.~J., Curry M.~B., de~Bruijn O.}:
\newblock The effects of visual information on users' mental models: An
  evaluation of pathfinder analysis as a measure of icon usability.
\newblock \emph{International Journal of Cognitive Ergonomics 5}, 1 (2001),
  59--84.

\bibitem[MHSG18]{mcinnes2018umap-software}
\textsc{McInnes L., Healy J., Saul N., Grossberger L.}:
\newblock Umap: Uniform manifold approximation and projection.
\newblock \emph{The Journal of Open Source Software 3}, 29 (2018), 861.

\bibitem[MI09]{mcdougall2009s}
\textsc{McDougall S., Isherwood S.}:
\newblock What’s in a name? the role of graphics, functions, and their
  interrelationships in icon identification.
\newblock \emph{Behavior research methods 41}, 2 (2009), 325--336.

\bibitem[PWS{\etalchar{*}}21]{Patashnik_2021_ICCV}
\textsc{Patashnik O., Wu Z., Shechtman E., Cohen-Or D., Lischinski D.}:
\newblock Styleclip: Text-driven manipulation of stylegan imagery.
\newblock In \emph{Proceedings of the IEEE/CVF International Conference on
  Computer Vision (ICCV)} (October 2021), pp.~2085--2094.

\bibitem[PXW18]{Peng:2018:AS}
\textsc{Peng M., Xing J., Wei L.-Y.}:
\newblock Autocomplete 3d sculpting.
\newblock \emph{ACM Trans. Graph. 37}, 4 (July 2018), 132:1--132:15.
\newblock URL: \url{http://doi.acm.org/10.1145/3197517.3201297}, \href
  {https://doi.org/10.1145/3197517.3201297}
  {\path{doi:10.1145/3197517.3201297}}.

\bibitem[RDF11]{rosenholtz2011predictions}
\textsc{Rosenholtz R., Dorai A., Freeman R.}:
\newblock Do predictions of visual perception aid design?
\newblock \emph{ACM Transactions on Applied Perception (TAP) 8}, 2 (2011),
  1--20.

\bibitem[RKH{\etalchar{*}}21]{radford2021learning}
\textsc{Radford A., Kim J.~W., Hallacy C., Ramesh A., Goh G., Agarwal S.,
  Sastry G., Askell A., Mishkin P., Clark J., et~al.}:
\newblock Learning transferable visual models from natural language
  supervision.
\newblock In \emph{International Conference on Machine Learning} (2021), PMLR,
  pp.~8748--8763.

\bibitem[SABAG{\etalchar{*}}05]{setlur2005semanticons}
\textsc{Setlur V., Albrecht-Buehler C., A.~Gooch A., Rossoff S., Gooch B.}:
\newblock Semanticons: Visual metaphors as file icons.
\newblock In \emph{Computer Graphics Forum} (2005), vol.~24, pp.~647--656.

\bibitem[SC21]{shen2021clipgen}
\textsc{Shen I.-C., Chen B.-Y.}:
\newblock Clipgen: A deep generative model for clipart vectorization and
  synthesis.
\newblock \emph{IEEE Transactions on Visualization and Computer Graphics 28},
  12 (2021), 4211--4224.

\bibitem[SL19]{swearngin2019modeling}
\textsc{Swearngin A., Li Y.}:
\newblock Modeling mobile interface tappability using crowdsourcing and deep
  learning.
\newblock In \emph{Proceedings of the 2019 CHI Conference on Human Factors in
  Computing Systems} (2019), pp.~1--11.

\bibitem[SLS{\etalchar{*}}21]{shen2020clipflip}
\textsc{Shen I.-C., Liu K.-H., Su L.-W., Wu Y.-T., Chen B.-Y.}:
\newblock Clipflip : Multi-view clipart design.
\newblock \emph{Computer Graphics Forum} (2021).
\newblock \href {https://doi.org/10.1111/cgf.14190}
  {\path{doi:10.1111/cgf.14190}}.

\bibitem[SM14]{setlur2014automatic}
\textsc{Setlur V., Mackinlay J.~D.}:
\newblock Automatic generation of semantic icon encodings for visualizations.
\newblock In \emph{Proceedings of the 32nd annual ACM conference on Human
  factors in computing systems} (2014), pp.~541--550.

\bibitem[SRS18]{spiliotopoulos2018comparative}
\textsc{Spiliotopoulos K., Rigou M., Sirmakessis S.}:
\newblock A comparative study of skeuomorphic and flat design from a ux
  perspective.
\newblock \emph{Multimodal Technologies and Interaction 2}, 2 (2018), 31.

\bibitem[SSII18]{SimoSerraSIGGRAPH2018}
\textsc{Simo-Serra E., Iizuka S., Ishikawa H.}:
\newblock {Real-Time Data-Driven Interactive Rough Sketch Inking}.
\newblock \emph{ACM Transactions on Graphics (SIGGRAPH) 37}, 4 (2018).

\bibitem[SZL{\etalchar{*}}21]{shen2021effects}
\textsc{Shen Z., Zhang L., Li R., Hou J., Liu C., Hu W.}:
\newblock The effects of color combinations, luminance contrast, and area ratio
  on icon visual search performance.
\newblock \emph{Displays 67} (2021), 101999.

\bibitem[Tak01]{takagi2001interactive}
\textsc{Takagi H.}:
\newblock Interactive evolutionary computation: Fusion of the capabilities of
  ec optimization and human evaluation.
\newblock \emph{Proceedings of the IEEE 89}, 9 (2001), 1275--1296.

\bibitem[UIM12]{umetani2012guided}
\textsc{Umetani N., Igarashi T., Mitra N.~J.}:
\newblock Guided exploration of physically valid shapes for furniture design.
\newblock \emph{ACM Trans. Graph. 31}, 4 (2012), 86--1.

\bibitem[UKSI14]{umetani2014airplanes}
\textsc{Umetani N., Koyama Y., Schmidt R., Igarashi T.}:
\newblock Pteromys: Interactive design and optimization of free-formed
  free-flight model airplanes.
\newblock \emph{ACM Trans. Graph. 33}, 4 (July 2014).
\newblock URL: \url{https://doi.org/10.1145/2601097.2601129}, \href
  {https://doi.org/10.1145/2601097.2601129}
  {\path{doi:10.1145/2601097.2601129}}.

\bibitem[VdOLV18]{van2018representation}
\textsc{Van~den Oord A., Li Y., Vinyals O.}:
\newblock Representation learning with contrastive predictive coding.
\newblock \emph{arXiv e-prints} (2018), arXiv--1807.

\bibitem[VSP{\etalchar{*}}17]{Vaswani2017trans}
\textsc{Vaswani A., Shazeer N., Parmar N., Uszkoreit J., Jones L., Gomez A.~N.,
  Kaiser L., Polosukhin I.}:
\newblock Attention is all you need.
\newblock URL: \url{https://arxiv.org/pdf/1706.03762.pdf}.

\bibitem[WMLB13]{warnock2013multiple}
\textsc{Warnock D., McGee-Lennon M., Brewster S.}:
\newblock Multiple notification modalities and older users.
\newblock In \emph{Proceedings of the SIGCHI Conference on Human Factors in
  Computing Systems} (2013), pp.~1091--1094.

\bibitem[XCW14]{Xing:2014:APR}
\textsc{Xing J., Chen H.-T., Wei L.-Y.}:
\newblock Autocomplete painting repetitions.
\newblock \emph{ACM Trans. Graph. 33}, 6 (Nov. 2014), 172:1--172:11.
\newblock URL: \url{http://doi.acm.org/10.1145/2661229.2661247}, \href
  {https://doi.org/10.1145/2661229.2661247}
  {\path{doi:10.1145/2661229.2661247}}.

\bibitem[XKG{\etalchar{*}}16]{Xing2016}
\textsc{Xing J., Kazi R.~H., Grossman T., Wei L.-Y., Stam J., Fitzmaurice G.}:
\newblock Energy-brushes: Interactive tools for illustrating stylized elemental
  dynamics.
\newblock In \emph{Proceedings of the 29th Annual Symposium on User Interface
  Software and Technology} (New York, NY, USA, 2016), UIST '16, Association for
  Computing Machinery, p.~755–766.
\newblock URL: \url{https://doi.org/10.1145/2984511.2984585}, \href
  {https://doi.org/10.1145/2984511.2984585}
  {\path{doi:10.1145/2984511.2984585}}.

\bibitem[ZKH{\etalchar{*}}20]{zhao2020iconate}
\textsc{Zhao N., Kim N.~W., Herman L.~M., Pfister H., Lau R.~W., Echevarria J.,
  Bylinskii Z.}:
\newblock Iconate: Automatic compound icon generation and ideation.
\newblock In \emph{Proceedings of the 2020 CHI Conference on Human Factors in
  Computing Systems} (2020), pp.~1--13.

\end{thebibliography}

\end{document}


\ifthenelse{\equal{\conf}{siggraph}}{
    \input{siggraph/siggraph_metadata}
    \begin{abstract}
Interface icons are prevalent in various digital applications.
Due to limited time and budgets, many designers rely on informal evaluation, which often results in poor usability icons.
In this paper, we propose a unique human-in-the-loop framework that allows our target users, \ie~novice and professional UI designers, to improve the usability of interface icons efficiently.
We formulate several usability criteria into a perceptual usability function and enable users to iteratively revise an icon set with an interactive design tool, EvIcon.
We take a large-scale pre-trained joint image-text embedding (CLIP)
and fine-tune it to embed icon visuals with icon tags in the same embedding space (IconCLIP).
During the revision process, our design tool provides two types of instant perceptual usability feedback.
First, we provide perceptual usability feedback modeled by deep learning models trained on IconCLIP embeddings and crowdsourced perceptual ratings.
Second, we use the embedding space of IconCLIP to assist users in improving icons' visual distinguishability among icons within the user-prepared icon set.
To provide the perceptual prediction, we compiled \textit{IconCEPT10K}, the first large-scale dataset of perceptual usability ratings over $10,000$ interface icons, by conducting a crowdsourcing study.
We demonstrated that our framework could benefit UI designers' interface icon revision process with a wide range of professional experience.
Moreover, the interface icons designed using our framework achieved better semantic distance and familiarity, verified by an additional online user study.
\end{abstract}

    \input{siggraph/siggraph_ccs_keyword}
}{}

\ifthenelse{\equal{\conf}{uist}}{
    \input{uist/uist_metadata}
    \begin{abstract}
Interface icons are prevalent in various digital applications.
Due to limited time and budgets, many designers rely on informal evaluation, which often results in poor usability icons.
In this paper, we propose a unique human-in-the-loop framework that allows our target users, \ie~novice and professional UI designers, to improve the usability of interface icons efficiently.
We formulate several usability criteria into a perceptual usability function and enable users to iteratively revise an icon set with an interactive design tool, EvIcon.
We take a large-scale pre-trained joint image-text embedding (CLIP)
and fine-tune it to embed icon visuals with icon tags in the same embedding space (IconCLIP).
During the revision process, our design tool provides two types of instant perceptual usability feedback.
First, we provide perceptual usability feedback modeled by deep learning models trained on IconCLIP embeddings and crowdsourced perceptual ratings.
Second, we use the embedding space of IconCLIP to assist users in improving icons' visual distinguishability among icons within the user-prepared icon set.
To provide the perceptual prediction, we compiled \textit{IconCEPT10K}, the first large-scale dataset of perceptual usability ratings over $10,000$ interface icons, by conducting a crowdsourcing study.
We demonstrated that our framework could benefit UI designers' interface icon revision process with a wide range of professional experience.
Moreover, the interface icons designed using our framework achieved better semantic distance and familiarity, verified by an additional online user study.
\end{abstract}

    \input{uist/uist_ccs_keyword}
}{}

\ifthenelse{\equal{\conf}{chi}}{
    \input{chi/chi_metadata}
    \begin{abstract}
Interface icons are prevalent in various digital applications.
Due to limited time and budgets, many designers rely on informal evaluation, which often results in poor usability icons.
In this paper, we propose a unique human-in-the-loop framework that allows our target users, \ie~novice and professional UI designers, to improve the usability of interface icons efficiently.
We formulate several usability criteria into a perceptual usability function and enable users to iteratively revise an icon set with an interactive design tool, EvIcon.
We take a large-scale pre-trained joint image-text embedding (CLIP)
and fine-tune it to embed icon visuals with icon tags in the same embedding space (IconCLIP).
During the revision process, our design tool provides two types of instant perceptual usability feedback.
First, we provide perceptual usability feedback modeled by deep learning models trained on IconCLIP embeddings and crowdsourced perceptual ratings.
Second, we use the embedding space of IconCLIP to assist users in improving icons' visual distinguishability among icons within the user-prepared icon set.
To provide the perceptual prediction, we compiled \textit{IconCEPT10K}, the first large-scale dataset of perceptual usability ratings over $10,000$ interface icons, by conducting a crowdsourcing study.
We demonstrated that our framework could benefit UI designers' interface icon revision process with a wide range of professional experience.
Moreover, the interface icons designed using our framework achieved better semantic distance and familiarity, verified by an additional online user study.
\end{abstract}

    \input{chi/chi_ccs_keyword}
}{}

\ifthenelse{\equal{\conf}{tvcg}}{
    \input{ieeetrans/ieeetrans_metadata}
}{}

\ifthenelse{\equal{\conf}{cvpr}}{
    \input{cvpr/cvpr_metadata}
}{}

\ifthenelse{\equal{\conf}{eccv}}{
    \input{eccv/eccv_metadata}
}{}

\ifthenelse{\equal{\conf}{pg}}{
\ifthenelse{\equal{\doc}{main}}{
\title{EvIcon: Designing High-Usability Icon with Human-in-the-loop Exploration and IconCLIP}
}
{
\title{Supplemental Material for EvIcon: Designing High-Usability Icon with Human-in-the-loop Exploration and IconCLIP}
}

\author[I-Chao Shen et al.]
{\parbox{\textwidth}{\centering I-Chao Shen$^{1}$ ~ Fu-Yin Cherng$^{2}$\thanks{Corresponding author} ~ Takeo Igarashi$^{3}$ ~ Wen-Chieh Lin$^{4}$ ~ Bing-Yu Chen$^{5}$
        }
        \\
{\parbox{\textwidth}{\centering $^1$ ichaoshen@g.ecc.u-tokyo.ac.jp
, The University of Tokyo, Japan \\
$^2$ fuyincherng@cs.ccu.edu.tw, National Chung Cheng University, Taiwan \\
$^3$ takeo@acm.org, The University of Tokyo, Japan \\
$^4$ wclin@cs.nycu.edu.tw, National Yang Ming Chiao Tung University, Taiwan \\
$^5$ robin@ntu.edu.tw, National Taiwan University, Taiwan \\
      }
}
}

}{}
\ifthenelse{\equal{\conf}{cgf}}{
\ifthenelse{\equal{\doc}{main}}{
\title{EvIcon: Designing High-Usability Icon with Human-in-the-loop Exploration and IconCLIP}
}
{
\title{Supplemental Material for EvIcon: Designing High-Usability Icon with Human-in-the-loop Exploration and IconCLIP}
}

\author[I-Chao Shen et al.]
{\parbox{\textwidth}{\centering I-Chao Shen$^{1}$ ~ Fu-Yin Cherng$^{2}$\thanks{Corresponding author} ~ Takeo Igarashi$^{3}$ ~ Wen-Chieh Lin$^{4}$ ~ Bing-Yu Chen$^{5}$
        }
        \\
{\parbox{\textwidth}{\centering $^1$ ichaoshen@g.ecc.u-tokyo.ac.jp
, The University of Tokyo, Japan \\
$^2$ fuyincherng@cs.ccu.edu.tw, National Chung Cheng University, Taiwan \\
$^3$ takeo@acm.org, The University of Tokyo, Japan \\
$^4$ wclin@cs.nycu.edu.tw, National Yang Ming Chiao Tung University, Taiwan \\
$^5$ robin@ntu.edu.tw, National Taiwan University, Taiwan \\
      }
}
}

}{}
\maketitle

\ifthenelse{\equal{\conf}{cvpr} \OR \equal{\conf}{eccv}}{
\begin{abstract}
Interface icons are prevalent in various digital applications.
Due to limited time and budgets, many designers rely on informal evaluation, which often results in poor usability icons.
In this paper, we propose a unique human-in-the-loop framework that allows our target users, \ie~novice and professional UI designers, to improve the usability of interface icons efficiently.
We formulate several usability criteria into a perceptual usability function and enable users to iteratively revise an icon set with an interactive design tool, EvIcon.
We take a large-scale pre-trained joint image-text embedding (CLIP)
and fine-tune it to embed icon visuals with icon tags in the same embedding space (IconCLIP).
During the revision process, our design tool provides two types of instant perceptual usability feedback.
First, we provide perceptual usability feedback modeled by deep learning models trained on IconCLIP embeddings and crowdsourced perceptual ratings.
Second, we use the embedding space of IconCLIP to assist users in improving icons' visual distinguishability among icons within the user-prepared icon set.
To provide the perceptual prediction, we compiled \textit{IconCEPT10K}, the first large-scale dataset of perceptual usability ratings over $10,000$ interface icons, by conducting a crowdsourcing study.
We demonstrated that our framework could benefit UI designers' interface icon revision process with a wide range of professional experience.
Moreover, the interface icons designed using our framework achieved better semantic distance and familiarity, verified by an additional online user study.
\end{abstract}

}{}


\ifthenelse{\equal{\conf}{tvcg}}{
\IEEEdisplaynontitleabstractindextext
\IEEEpeerreviewmaketitle
}{}

\section{Icon data curation and AMT crowdsource rating collection}
\label{sec:supp_amt_collection}

\subsection{Rating distribution}

\begin{figure}
\centering
  \includegraphics[width=\linewidth]{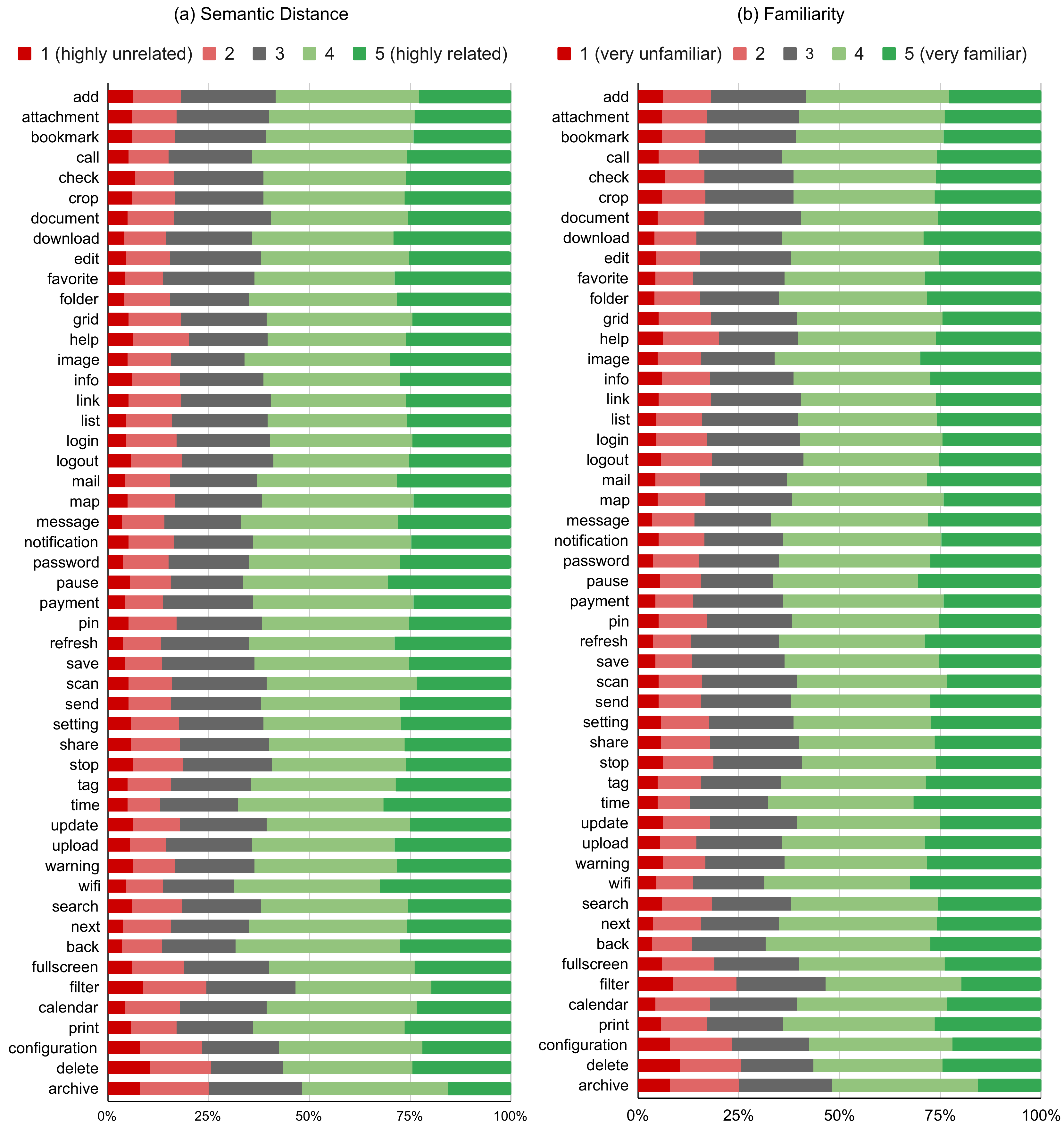}
  \caption{
  The percentage of the (a) semantic distance and (b) familiarity levels for each tag in the crowdsourced labeling dataset.}
\label{fig:per_dist}
\end{figure}

\Cref{fig:per_dist} illustrates the percentage of rating level (1-5) in each base tag for semantic distance and familiarity. We can see that the distribution of rating levels varies for each tag, and level ``4'' occupied the largest proportion among other levels. 

\subsection{Crowdsource task process}
\label{sec:task_process}
Before rating the icons, crowdworkers were asked to report the demographic questions, including age, sex, and occupation.
Then, they were asked to read the instructions about (i) the definition of semantic distance and familiarity of icons and (ii) the conditions of rejections.
In the rating task, crowdworkers were first asked to rate the familiarity of the presented tag on a scale from 1 (very unfamiliar) to 5 (very familiar).
Next, five icons of this function were displayed and crowdworkers were asked to rate the semantic distance of each icon by \textit{``How semantically related is this icon to the tag from 1 (highly unrelated) to 5 (highly related)?''}~\cite{mcdougall1999measuring,isherwood2007icon}, and the familiarity of each icon by \textit{``How familiar is this icon to you from 1 (very unfamiliar) to 5 (very familiar)?''}~\cite{mcdougall1999measuring}.
We inserted one repeated icon as a sanity check question among five icons to detect whether the crowdworkers provided contradictory ratings to the same icon.  
The workers would rate five tags in each assignment (\ie~25 icons in total), and the order of the tags and icons was randomized.
The average completion time was 8 minutes.

Those workers who meet any of the following criteria would be treated as outliers and removed from the rating dataset: (i) contradictory answers for the sanity check questions, (ii) giving the same rating to all icons, and (iii) failing to rate all icons. 
We also removed eight workers who reported their ages were less than five. As a result, there are 183 outliers removed from our dataset. The rejection rate is 6.8\%.

\section{The original Icon and Revised Icons in Evaluation}
\begin{figure*}[h!]
\centering
  \includegraphics[width=0.8\linewidth]{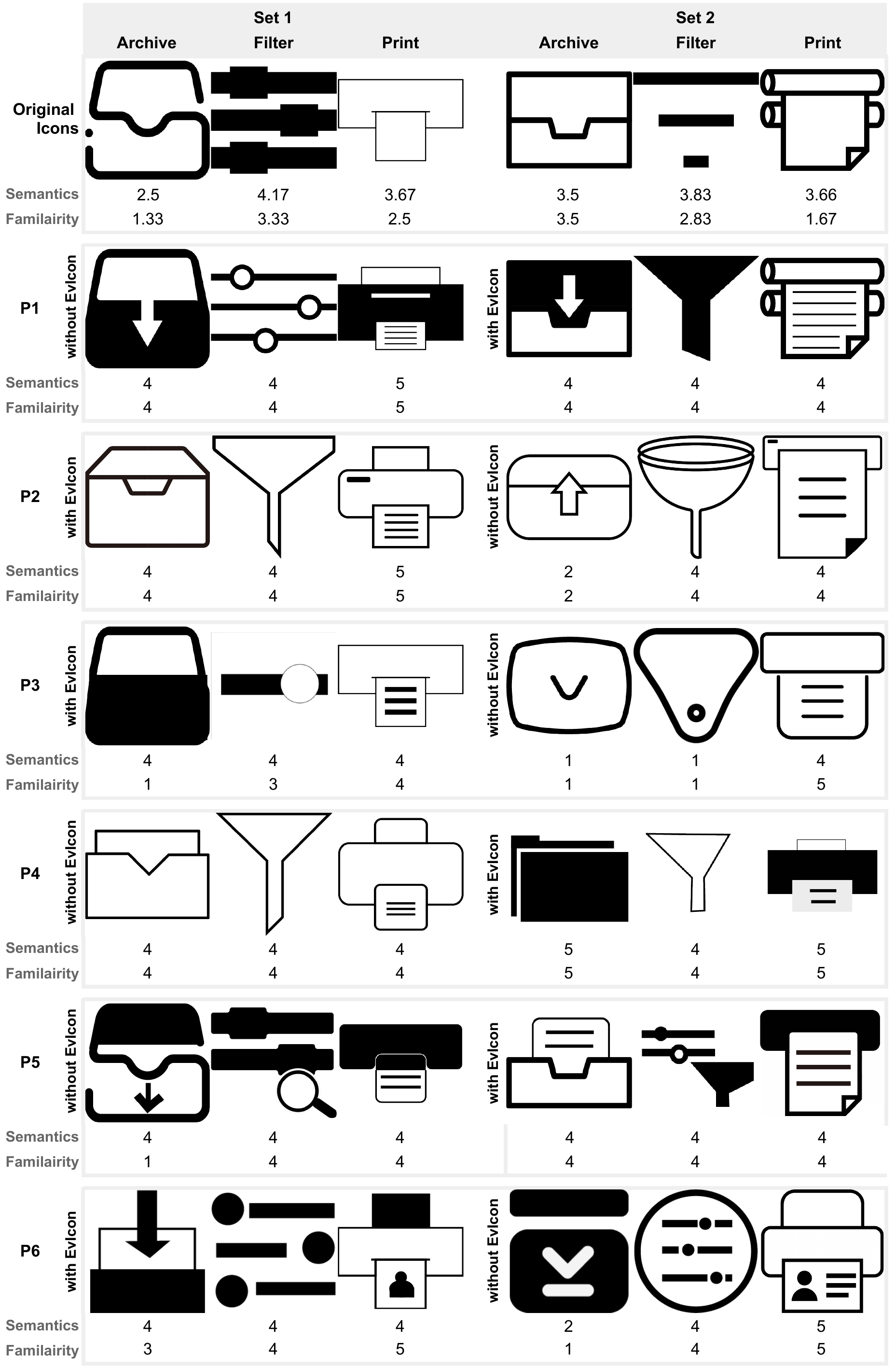}
  \caption{The original icons used in our evaluation and the revised icons by six UI designers. For each icon revised by a designer, we used the mode score provided by crowdworkers for the semantics (semantic distance) and familiarity. The usability scores of the original icons were calculated by averaging the scores provided by the crowdworkers for each designer separately.
}
\label{fig:design_outcomes}
\end{figure*}

In \Cref{fig:design_outcomes}, we show all revised icons in our study with UI designers.

\ifthenelse{\equal{\conf}{siggraph} \OR \equal{\conf}{uist} \OR \equal{\conf}{chi}}{
    \bibliographystyle{ACM-Reference-Format}
    \bibliography{paper}
}{}

\ifthenelse{\equal{\conf}{cvpr}}{
    \bibliographystyle{ieee_fullname}
    \bibliography{paper}
}{}

\ifthenelse{\equal{\conf}{eccv}}{
    \bibliographystyle{splncs04}
    \bibliography{paper}
}{}

\clearpage
\ifthenelse{\equal{\conf}{pg} \OR \equal{\conf}{eg} \OR \equal{\conf}{cgf}}{
    \bibliographystyle{eg-alpha-doi} 
    \bibliography{paper}       
}{}
